\documentclass[prd]{revtex4}
\usepackage{epsfig}
\usepackage{mathrsfs}
\usepackage{amsmath}
\usepackage{amssymb}
\usepackage{color}
\usepackage{graphicx}
\usepackage{comment}

\begin{document}

\title{The spectral shapes of Galactic gamma--ray sources} 

\author{Paolo Lipari}
\email{paolo.lipari@roma1.infn.it}

\affiliation{INFN, Sezione Roma ``Sapienza'',
 piazzale A.Moro 2, 00185 Roma, Italy}

\affiliation{
 IHEP, Key Laboratory of Particle Astrophysics,
 Chinese Academy of Sciences, Beijing, China}

\author{Silvia Vernetto}
\email{vernetto@to.infn.it}

\affiliation{
INAF, Osservatorio Astrofisico di Torino,
 via P.Giuria 1, 10125 Torino, Italy}

\affiliation{
 INFN, Sezione Torino,
 via P.Giuria 1, 10125 Torino, Italy}

\begin{abstract}
Recent observations by ground--based gamma--ray telescopes
have led to the publication of catalogs 
listing sources observed in the TeV and PeV energy ranges.
Photons of such high energy are strongly absorbed during
propagation over extragalactic distances,
and the catalogs are dominated by Galactic sources.
Of particular interest are the observations of 
the LHAASO telescope, which cover
a very broad energy range (from 1 to 10$^3$~TeV)
and show that the spectra
of all Galactic gamma--ray sources are curved,
with significantly different slopes 
below and above $E \sim 30$~TeV.
The cumulative spectrum obtained by summing the contributions
of Galactic individual sources
has a spectral shape that gradually softens with energy,
with a slope that increases from a value of order 2.2
at $E \simeq 1$~TeV, to 2.5 at 30~TeV, and $\simeq 3.4$ at 100~TeV.
It is remarkable that the smooth variation in the
shape of the cumulative spectrum is obtained from the sum of
contributions that have a wide range of shapes.
Understanding the origin of the spectral shapes of the
Galactic gamma--ray sources
is a crucial challenge for high energy astrophysics.
\end{abstract}

\maketitle

% \tableofcontents
%\clearpage

\section{Introduction}
\label{sec:introduction}

Over the past few decades, space--based \cite{Fermi-LAT:2019yla,Fermi-LAT-4FGL-dr4}
and ground--based
\cite{HAWC:2020hrt,HESS:2018pbp,LHAASO:2023rpg}
telescopes
have produced a wealth of new results measuring gamma--rays
in a wide range of energies spanning over seven orders of magnitude
from 100~MeV to more than 1~PeV.
The observations have revealed the existence
of a large number of point--like and quasi--point--like sources,
belonging to different classes of astrophysical objects,
and a diffuse flux that varies smoothly with the direction $\Omega$.
The diffuse flux is formed by three components:
\begin{equation}
 \phi_{\rm diff} (E, \Omega) = \phi_{\rm ism} (E, \Omega) +
 \phi_{\rm sources}^{\rm unres.} (E, \Omega) + \phi_{\rm extra} (E) ~.
\end{equation}
The third component on the right hand side of this equation
is of extragalactic origin \cite{Fermi-LAT:2014ryh}, it is,
to a good approximation isotropic, and it 
 will not be discussed in this paper, while the first two are both Galactic.
The first component is formed by the photons
created by cosmic rays (CR) propagating in the interstellar medium 
of the Milky Way and interacting with gas and radiation fields
\cite{Fermi-LAT:2014ryh,Fermi-LAT:2016zaq,TibetASgamma:2021tpz,LHAASO:2023gne,HAWC:2023wdq},
the second one is due to the emission of the ensemble of unresolved sources
\cite{Lipari:2018gzn,Steppa:2020qwe,Cataldo:2020qla,Luque:2022buq}.
It should be noted that the flux of the unresolved sources
is not a fundamental quantity, since it depends on the sensitivity
of the observations, and in principle it becomes negligibly small for
a very high sensitivities.

The (truly diffuse) interstellar emission flux and the
unresolved source flux have different
dependences on energy and direction and can therefore, at least in principle,
be separated. However, the form of these dependences,
which actually encode very important information, are not known a priori,
and must be determined from the observations.
In this paper we want to address the question of the spectral shape
of the unresolved source flux. This shape will be estimated
from a study of the ensemble of Galactic sources that are resolved.

In general the spectral shape of the unresolved sources flux
will be a function of the direction $\Omega$, because looking in different
directions a telescope observes different parts of the Milky Way
that can contain different populations of sources.
However, in this paper, motivated by the search for simplicity,
we will assume that the spectral shape of the unresolved sources flux
is independent from direction and, at least to a first approximation,
also equal to the shape of the sum of the spectra of the resolved sources.
This common spectral shape will then also apply to
the total Galactic source emission $Q_{\rm MW}^{\rm sources}(E)$ which is obtained
by summing with equal weight, regardless of position and absolute luminosity,
the emissions of all sources in the Galaxy.
This would be the observable quantity for an extragalactic oberver located
at a large distance from the Milky Way, that appears as a quasi point--like object.
These assumptions can be stated as:
\begin{equation}
 \phi_{\rm sources}^{\rm unres.} (E, \Omega) \propto \left ( \sum_j^{\rm resol.} \phi_j (E) \right )
\propto Q_{\rm MW}^{\rm sources} (E) = \left (\sum_j^{\rm all} q_j (E) \right ) 
\label{eq:sources_factorization}
\end{equation}
where the first (second) summation is over resolved (all) Galactic sources.

It is possible, and indeed likely, that the assumptions
in Eq.(\ref{eq:sources_factorization}) are not exactly valid, for several reasons:
(i) the populations of sources near the center and in the periphery of the Galaxy
are likely not to be identical; (ii) resolved sources are on average
less distant, brighter and perhaps less extended
than unresolved ones, and therefore may also
emit different spectral shapes; (iii) the summation of fluxes (rather than emissions) 
gives more weight ($\propto d^{-2}$) to closer objects.
It is desirable, and indeed necessary to address these issues, 
but in this paper we will use Eq.(\ref{eq:sources_factorization})
as a simple, natural first approximation for estimating the shape
of the total Galactic source spectrum and of the unresolved source spectrum.
It is important to note that the validity of the assumptions
in Eq.(\ref{eq:sources_factorization}) can be tested
by comparing the spectra of resolved sources of different type
and in different regions of the sky.

A main motivation for this study of the spectral shape of the
source flux is that a good description of this shape
can help to separate the contributions of unresolved sources
and interstellar emission in the observed diffuse flux.
On the other hand, the measurement of the total source flux
and of its spectral shape is of considerable interest in itself.

The most interesting result that emerges from our study
is that the total Galactic flux of the gamma--ray sources
has a smooth shape, with only few well defined features,
and it is formed by the sum of contributions from
individual sources that emit spectra with a wide range of shapes.
Therefore, the assumption made in many studies that
all (or most) of the Galactic gamma--ray sources emit spectra
with the same shape is not valid.

Our discussion on the gamma--ray fluxes is also relevant
to the interpretation of neutrino observations
\cite{IceCube:2013low,IceCube:2014stg,IceCube:2019cia,IceCube:2023ame}.
This is because the gamma--ray and neutrino emissions are closely related.
The high energy neutrino flux is formed by the same
components as the gamma--ray flux, and also includes
an ensemble of point--like sources and a Galactic diffuse flux
generated by CR particles during propagation in interstellar space.
The relation between the neutrino and gamma--ray fluxes is determined
by the relative importance of the hadronic and leptonic mechanisms
in gamma--ray emission, since the photon emission generated by the 
former is accompanied by a $\nu$ emission of the same size and shape.

The paper is organized as follows.
In the next section we will introduce the four gamma--ray source catalogs,
obtained by the Fermi--LAT, HESS, HAWC and LHAASO telescopes,
that will be used in our study,
and discuss the functional forms that are used in the catalogs
to fit the spectra of individual sources.
Section~\ref{sec:fermi} discusses the sources observed by the Fermi--LAT
telescope in the energy range 0.1--10$^3$~GeV. These sources are partly Galactic
and partly extragalactic, but their distribution over the celestial sphere
allows to identify the Galactic ones.
Section~\ref{sec:tev_sources} discusses the observations at higher energies
($E \gtrsim 1$~TeV) that are only possible with ground--based telescope
and therefore limited to only a part of the celestial sphere.
In this range the flux from extragalactic sources is strongly suppressed
due to the absorption of the gamma--rays during propagation.
Of particular importance are the recent observations with the LHAASO telescope,
that can measure the spectra in a wide interval extending
from~1 to $10^3$~TeV.
The last section contains a brief summary and a final discussion.

\section{Catalogs of gamma--ray sources}
\label{sec:catalogs}
In this work we will consider the gamma--ray sources listed in four catalogs.
\begin{enumerate}
\item The fourth general catalog of Fermi--LAT, using the 
4th data release (4FGL-DR4)
\cite{Fermi-LAT:2019yla,Fermi-LAT-4FGL-dr4}.
This catalog contains 7195 gamma--ray
sources observed during 14 years of data acquisition
in the energy range from 50~MeV to 1~TeV with
a sensitivity that, to a good approximation, is uniform 
over the entire celestial sphere.

\item The HESS Galactic Plane Survey (HGPS) catalog \cite{HESS:2018pbp}
 obtained by the High Energy Stereoscopic System (HESS),
 an array of imaging atmospheric Cherenkov telescopes located
 in Namibia, at geographic latitude $-23.27^\circ$.
 The catalog lists 78 sources observed in the energy range 0.2--100~TeV,
 during a scan (2700 hours of data taking) of the Galactic plane 
 in the region of Galactic latitude
 $|b| < 3^\circ$ and
 longitude $ -110^\circ \le \ell \le +65^\circ$.

\item 
 The third HAWC Catalog (3HAWC) \cite{HAWC:2020hrt} obtained by the
 High--Altitude Water Cherenkov observatory, located
 in Mexico at the geographic latitude
 $\lambda = 19.0^\circ$N and altitude 4100~m.
 The catalog lists 65 sources, observed
 in the 0.1--100~TeV energy range during 1523 days of data taking
 (between November 2014 and June 2019).

\item 
 The first LHAASO catalog (1LHAASO) \cite{LHAASO:2023rpg}, obtained by the
 Large High--Altitude Air Shower Observatory (LHAASO), located
 in China at latitude $\lambda = 29.36^\circ$N and altitude 4400~m.
 The telescope is formed by two arrays:
 the water Cherenkov detector array (WCDA),
 sensitive in the energy range (1--30)~TeV,
 and the Kilometer Squared Array (KM2A), sensitive in the range from about 25~TeV
 to more than 1~PeV.
 % The LHAASO observatory contains also
 %a third array of wide--field--of--view Cherenkov/fluorescence
 %telescopes (WFCTA) but this instrument has not been used in the
 %construction of the catalog.
 The catalog is composed by two parts
 which list the sources resolved by the two arrays
 during 508 days of data acquisition for WCDA and 933 days for KM2A.
 The catalog contains a total of 90 sources,
 69 are observed in WCDA and 75 in KM2A, with 54 sources seen by
 both instruments. Out of the 75 sources observed by KM2A,
 44 have measurable fluxes also for $E > 100$~TeV.
\end{enumerate}
The HAWC and LHAASO telescopes are
Extensive Air Shower (EAS) detectors and therefore
can observe a limited range of celestial declination 
with an exposure per unit time determined by their geographic latitude.

\begin{figure}
\begin{center}
\includegraphics[width=12.9cm]{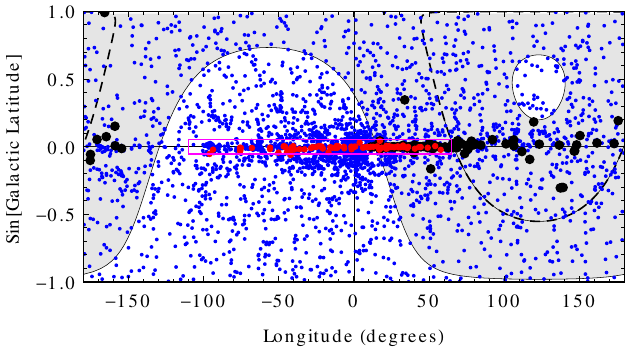}

\vspace{1.0cm}
\includegraphics[width=12.9cm]{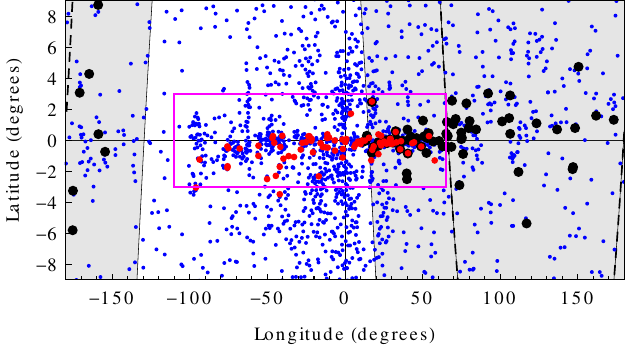}
\end{center}
\caption {\footnotesize
 Positions in the sky of sources in the different gamma--ray catalogs.
 The small blue points are sources in the 4FGL catalog (excluding extragalactic objects).
 The medium size red points are sources in the HGPS catalog. The large black points are
 sources in the LHAASO catalog (also excluding extragalactic sources).
 A red box indicates the sky region of the HGPS survey.
 The shaded area is the sky region visible to LHAASO
 (assuming that observations are possible
 when a source has zenith angle $\theta < 50^\circ$). The dashed lines indicate the points
 on the celestial sphere that pass through the zenith at the LHAASO latitude.
 The bottom panel shows an enlargement of the sky region around the
 Galactic equator.
\label{fig:catalog1} }
\end{figure}

The positions in sky of the sources in the 4FGL, HGPS and 1LHAASO catalogs
are shown in Fig.~\ref{fig:catalog1}. The figure also shows the
region of the HGPS survey and the region observable by the LHAASO telescope.

\subsection{Spectral shapes of the gamma--ray sources}
\label{sec:spectral-shapes} 
To describe gamma--ray spectra spectra it is 
useful to define two quantities,
the index (or slope) $\alpha (E)$: 
\begin{equation}
 \alpha (E) = - \frac{d \ln \phi(E)}{d \ln E}
 % = -\frac{E}{\phi(E)} \, \frac{d \phi(E)}{d\ln (E)}
 ~, 
\end{equation}
and the curvature $\beta (E)$: 
\begin{equation}
 \beta(E) = - \frac{1}{2} \frac{d^2 \ln \phi(E)}{(d \ln E)^2} ~.
\end{equation}
%silvia [Add explanation]
Both quantities are in general energy dependent.
According to the definitions given above,
a larger (smaller) spectral index corresponds to
a softer (harder) spectrum, and a positive (negative) curvature parameter corresponds to
a downward concave (convex) spectrum that becomes softer (harder) with increasing energy.

The gamma--ray catalogs discussed in this work
include fits to the energy spectra of the observed sources based on some simple functional forms.
The Fermi--LAT Collaboration uses three functional forms to fit the spectra:
(i) power--law (PL), (ii) log--parabola (LP), 
and (iii) power--law with cutoff (CUT).
For each source the 4FGL catalog gives a ``pivot energy'' $E_0 $ where the spectrum is well measured,
and where the correlation between the fit parameters is smallest.
For all three functional forms, two of the fit parameters
are the value of differential flux $K_0 = \phi(E_0)$,
and the spectral index $\alpha_0 = \alpha(E_0)$ at the pivot energy $E_0$.

The power--law fit has the scale invariant form:
\begin{equation}
\phi(E) = K_0 ~\left (\frac{E}{E_0} \right)^{-\alpha_0} 
\label{eq:PL} 
\end{equation}
with an energy independent spectral index, and a 
curvature that is identically zero. 

The log--parabola spectral shape is parameterized in the catalog with the form: 
\begin{equation}
\phi(E) = K_0\; \left ( \frac{E}{E_0} \right )^{-(\alpha_0 + \beta \, \ln E/E_0)} 
\label{eq:log_parabola}
\end{equation}
where $K_0$ and $\alpha_0$ are again the absolute value of the flux
and the spectral index at the pivot energy $E_0$,
and $\beta$ is a constant curvature parameter. 
The spectral index depends linearly on $\ln E$:
\begin{equation}
 \alpha(E) = \alpha_0 + 2 \beta \, \ln \frac{E}{E_0} ~.
 \label{eq:alpha_log--parabola}
\end{equation}
The log--parabola spectral shape is {\em not}
scale invariant, and in the following will be re--parameterized with the form:
\begin{equation}
\phi(E) = K^* \; \left ( \frac{E}{E^*} \right )^{-(2 + \beta \, \ln E/E^*)} 
\label{eq:log_parabola1}
\end{equation}
introducing the characteristic energy $E^*$, defined as 
the energy where the spectrum
has slope $\alpha (E^*) = 2$, or equivalently as the energy where
the spectral energy distribution (SED) $E^2 \, \phi(E)$ has its maximum.
The quantity $K^* = \phi(E^*)$ is the flux at the characteristic energy.
The spectral index can be expressed in terms of the characteristic energy as: 
\begin{equation}
 \alpha(E) = 2 + 2 \beta \, \ln \frac{E}{E^*} ~.
 \label{eq:alpha_log--parabola1}
\end{equation}
The parameters $E^*$ and $K^*$ introduced in Eq.~(\ref{eq:log_parabola1})
are related to $\alpha_0$ and $K_0$ by:
\begin{equation}
E^* = E_0 \, e^{(2-\alpha_0)/(2 \beta)} ~,
\end{equation}
\begin{equation}
K^* = K_0 \, e^{(2-\alpha_0)/\beta} ~.
\end{equation}
It is interesting to note \cite{Lipari:2020szc} that, for a positive
curvature ($\beta > 0$), the log--parabola form can in principle be extended
to all energies because, in contrast to the power--law form which
implies a divergent energy emission
at high (for a spectra index $\alpha \le 2$) and/or low energies
(for $\alpha \ge 2$), integrating over all $E$
one obtains finite emissions, both in number of photons and in energy,
\begin{equation}
\Phi_{\rm tot} = \int_0^\infty dE~\phi(E) = K^* \, E^* \; \sqrt{\frac{\pi}{\beta}} \, e^{1/(4 \beta)}~,
\end{equation}
\begin{equation}
F_{\rm tot} = \int_0^\infty dE~E~\phi(E) = K^* \, (E^*)^2 \; \sqrt{\frac{\pi}{\beta}}~.
\end{equation}

The third spectral shape used in the Fermi--LAT catalog
is a power--law with a high energy cutoff:
\begin{equation}
 \phi(E) = K_0 \, \left (\frac{E}{E_0} \right )^{\frac{d}{b} -\alpha_0}
 ~\exp \left [ -\frac{d}{b^2} \; \left (1 - \left( \frac{E}{E_0}\right)^b
 \right )\right ]~.
\end{equation}
This form depends on the four parameters $K_0$, $\alpha_0$, $d$ and $b$.
As in the previous cases $K_0$ and $\alpha_0$ are the flux and spectral index
at the pivot energy $E_0$, while the adimensional parameters $d$ and $b$
describe the shape of the high energy cutoff.
Also in this case the spectrum is obviously not scale invariant
as the existence of the cutoff defines a characteristic energy $E_{\rm cut}$.
In the following we will use a more physically transparent parameterization
of this spectral shape that makes use of the cutoff energy: 
\begin{equation}
 \phi(E) = K_{\rm cut} \, e^{1/b} \, \left (\frac{E}{E_{\rm cut}} \right )^{-\alpha_{\rm low}}
 ~\exp \left [ -\frac{1}{b} \, \left( \frac{E}{E_{\rm cut} }\right)^b \right ] ~.
\label{eq:cutoff1}
\end{equation}
The slope and curvature of the spectrum can then be expressed as:
\begin{equation}
\alpha(E) = \alpha_{\rm low} + \left (\frac{E}{E_{\rm cut}} \right )^b ~,
\label{eq:alpha_cutoff}
\end{equation}
\begin{equation}
\beta(E) = \frac{b}{2} \, \left (\frac{E}{E_{\rm cut}} \right )^b
\end{equation}
with $\alpha_{\rm low}$ the spectral index for low energies (that is for $E \ll E_{\rm cut}$).
The cutoff energy $E_{\rm cut}$ is defined as the energy 
where the spectral index takes the value
$\alpha (E_{\rm cut}) = \alpha_{\rm low} + 1$, and $K_{\rm cut}$ is the value of the flux at the cutoff energy.
The adimensional parameter $b$
describes how fast the cutoff develops.
%and its meaning is manifest in Eq.~(\ref{eq:alpha_cutoff}).
For $b=1$ the cutoff
is a simple exponential, for $b > 0$ ($b< 0$) the cutoff is super
(sub)--exponential.
%It should be noted that the parameter $b$ in most cases (for 241 sources,
%that is 87\% of the total) is kept fixed in the fitting procedure
%with value $b=2/3$.

The parameters $\alpha_{\rm low}$ and $E_{\rm cut}$ introduced in Eq.~(\ref{eq:cutoff1})
are related to those used in the Fermi--LAT catalog by:
\begin{equation}
\alpha_{\rm low} = \alpha_0 - \frac{d}{b}
\end{equation}
and
\begin{equation}
E_{\rm cut} = E_0 \, \left (\frac{b}{d} \right )^{1/b} ~.
\end{equation}

The sources in the HGPS catalog are fitted either with a simple power--law
spectrum (as in Eq.~\ref{eq:PL}) or with a power--law with
an exponential cutoff of form:
\begin{equation}
\phi(E) = K ~\left (\frac{E}{E_0} \right )^{-\alpha_{\rm low}} ~e^{-E/E_{\rm cut}} ~.
\label{eq:PL_exp}
\end{equation}
This is a special case of the cutoff form used by Fermi
with $b=1$ and expressed in term of a (source dependent) pivot energy $E_0$.

The HAWC and LHAASO catalogs give a simple power--law fit
for each source. In both cases the fit parameters are the
(constant) spectral index and the flux at a reference energy
(equal for all sources).

\subsection{The shape of the cumulative spectrum}
\label{sec:cumulative_spectrum}
In this work we will consider the cumulative spectrum obtained by summing
the contributions of all objects in one set of sources:
\begin{equation}
 \phi_{\rm cum} (E) = \sum_j \phi_j (E) 
\end{equation}
The spectral index and curvature of the
cumulative flux can be expressed in terms of the same
parameters for the individual sources \cite{Lipari:2020szc}.
For the spectral index one has:
\begin{equation}
 \alpha_{\rm cum} (E) = \sum_j w_j (E) ~\alpha_j (E)
 = \langle \alpha (E) \rangle 
\label{eq:alpha_average}
\end{equation}
where $\langle \alpha (E) \rangle$ is the average of the
spectral index of the sources contributing to the total
calculated using a weight proportional to the flux;
$w_j (E)= \phi_j(E)/\phi_{\rm cum} (E)$.

The curvature of the cumulative specrum can be calculated as:
\begin{equation}
 \beta_{\rm cum} (E) = \langle \beta (E) \rangle - \frac{1}{2} \sigma_\alpha^2 (E) ~,
\label{eq:beta_sum}
\end{equation}
where $\langle \beta(E)\rangle$ is the average of the curvature of
the individual sources, and $\sigma_\alpha(E)$ is the r.m.s. of the spectral index distribution
\begin{equation}
\langle \beta(E) \rangle = \sum_j w_j (E) ~\beta_j (E) ~,
\end{equation}
and
\begin{equation}
\sigma_\alpha^2 (E) = \sum_j w_j (E) ~\alpha_j^2 (E) - \langle \alpha(E) \rangle^2 ~.
\end{equation}
Note that the curvature of the cumulative spectrum is not the average of the curvatures of
the individual sources, and also 
depends on the width of the spectral index distribution of these ources. 

An important consequence of Eq.~(\ref{eq:beta_sum}) is that an
ensemble of power--law spectra with different spectral indices
will always have a negative curvature and gradually harden.
This is because the term $\langle \beta(E)\rangle$ in Eq.~(\ref{eq:beta_sum})
vanishes (because all $\beta_j$ are zero)
and $\sigma^2_\alpha$ is positive by construction.
This result simply reflects the fact that the soft (large spectral index)
sources dominate the flux at lower energies, and the hard
(small spectral index) sources dominate at higher energies.
A cumulative spectrum formed by components of
different shape can have a positive curvature, and be gradually softening,
only if the spectra of the individual components are curved
and are (at least on average) softening \cite{Lipari:2020szc}.

\section{The Fermi-LAT Galactic sources} 
\label{sec:fermi}
The Fermi--LAT 4FGL--DR4 catalog \cite{Fermi-LAT:2019yla,Fermi-LAT-4FGL-dr4}
contains a total of 7195 sources,
of which 4618 (64.2\% of the total) are associated to a known astrophysical object.
The association allows to establish the nature of a source
and to determine if it is Galactic or extragalactic.
For the remaining 2577 sources (35.8\% of the total) there is no such association
and therefore their nature remains ambiguous.

\begin{figure}
\begin{center}
\includegraphics[width=15.0cm]{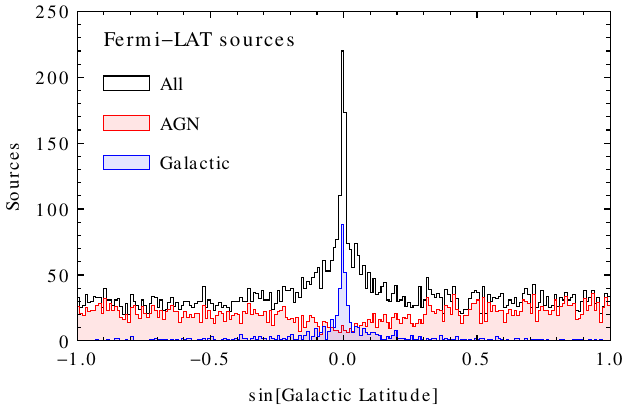}
\end{center}
\caption {\footnotesize
 Galactic latitude distribution of the sources in the Fermi--LAT 4FGL-DR4 catalog.
 The three histograms show the $\sin b$ distribution for all
 sources, AGN sources, and sources identified as Galactic.
\label{fig:fermi_slat} }
\end{figure}

The latitude distribution of the Fermi--LAT sources (see Fig.~\ref{fig:fermi_slat})
clearly shows the presence of an isotropic (flat in $\sin b$) extragalactic component
and of a Galactic component that forms a narrow peak centered at the equator
($b \simeq 0$) with a width of a few degrees,
that extends to $|b| \simeq 30^\circ$ and beyond.
Assuming that the extragalactic component is isotropic, and taking into
account for the angular dependence of the minimum flux to resolve a source
(that is larger in the sky regions where the background of the diffuse Galactic emission
is higher), we have estimated that the 4FGL catalog contains approximately
$2100 \pm 100$ Galactic sources and $5100 \pm 100$ extragalactic sources.

The catalog divides the sources into a total of 25 different classes
(with most of these classes consisting of two subclasses for
objects that are ``identified'' and those that are only ``associated'').
Twelve classes contain Galactic objects (with a special class for 
the source at the Galactic center), eleven contain extragalactic objects,
one class contains sources associated with objects of unknown nature,
and another is for objects that do not have an association.

A simplified summary of this source classification
is given in Table~\ref{tab:table_fermi_classes}.
The number of sources in the different classes are
given for the whole sky and for the Galactic disk region ($|b|< 3^\circ$).
The table also gives the number of 
sources fitted with the three functional forms discussed in
Sec.~\ref{sec:spectral-shapes} (power--law, log--parabola and cutoff).

%%%%%%%%%%%%%%%%%%%%%%%%%%%%%%%%%%%%%%%%%%%%%%%%%%%%%%%%%%%%%%%%%%%%%%%%%%%%%%%%%%%%%%%%%
\begin{table}[hbt]
\caption{\footnotesize
Classification of the gamma--ray sources in the
Fermi-LAT catalog 4FGL--DR4.
The left--hand part of the table includes all sources in the catalog,
while the right--hand part includes only sources in the disk region
($|b| < 3^\circ$). The number of sources fitted with the power--law 
(PL), log--parabola (LP) and cutoff (Cut.) forms is also
given. A horizontal line separates Galactic and extragalactic classes of sources.
\label{tab:table_fermi_classes}}
\renewcommand{\arraystretch}{1.2}

\vspace{0.25 cm}
 \begin{tabular}{ | l || r | r | r | r || r | r | r | r |}
 \hline
Class & \multicolumn{4}{c ||} {All sky} 
 & \multicolumn{4}{ c |} {$|b| \le 3^\circ$} \\
 \hline
 & & PL & LP & Cut. 
 & & PL & LP & Cut. \\
 \hline
Galactic Center	&	1	&	0	&	1	&	0	&	1	&	0	&	1	&	0	\\
Pulsars (PSR) 	&	141	&	6	&	3	&	132	&	108	&	3	&	1	&	104	\\
Millisecond pulsars (MSP) ~~&	179	&	6	&	36	&	137	&	14	&	0	&	4	&	10	\\
Supernova Remnants (SNR) &	45	&	14	&	31	&	0	&	35	&	10	&	25	&	0	\\
Pulsar Wind Nebulae (PWN) ~~&	22	&	15	&	7	&	0	&	17	&	12	&	5	&	0	\\
Composite SNR/PWN (SPP) ~~&	124	&	43	&	81	&	0	&	91	&	28	&	63	&	0	\\
Globular Clusters (GLC) ~~ &	34	&	5	&	29	&	0	&	6	&	0	&	6	&	0	\\
Star Formation Regions	 (SFR) ~~ &	6	&	3	&	3	&	0	&	4	&	3	&	1	&	0 \\
Binary systems (BIN)	&	30	&	9	&	21	&	0	&	10	&	3	&	7	&	0	\\
Novae	&	8	&	4	&	4	&	0	&	3	&	1	&	2	&	0	\\
\hline
Total Galactic sources & 590 & 105 & 216
& 
269
& 
289
& 
60
& 
115
& 
114
\\
\hline
Galaxies (GAL)	&	6	&	2	&	3	&	1	&	0	&	0	&	0	&	0	\\
Starburst galaxies (SBG)	&	8	&	6	&	2	&	0	&	0	&	0	&	0	&	0	\\
Seyfert galaxies	&	8	&	2	&	6	&	0	&	0	&	0	&	0	&	0	\\
Active Galactic Nuclei (AGN)	&	4006	&	2429	&	1571	&	6	&	90	&	37	&	53	&	0	\\
\hline
Total Extragalactic sources & 4028
& 
2439
& 
1582
& 
7
& 
90
& 
37
& 
53
& 
0
\\
\hline
No-ID	&	2577	&	1299	&	1278	&	0	&	628	&	196	&	432	&	0	\\
\hline
All sources	&	7195	&	3843	&	3076	&	276	&	1007	&	293	&	600	&	114	\\ \hline
\end{tabular}
\end{table}
%%%%%%%%%%%%%%%%%%%%%%%%%%%%%%%%%%%%%%%%%%%%%%%%%%%%%%%%%%%%%%%%%%%%%%%%%%%%%%%%%%%%%%%%%%%%%%%
In preparing this table we have made the following simplifications:
(a) the distinction between identified and associated
sources has been ignored,
(b) eight different types of Active Galactic Nuclei have been grouped
into a single class (AGN), leaving Seyfert galaxies as a separate class,
(c) three types of binary star systems have been grouped into a single class (BIN),
(d) sources with no associations and sources associated with
an object of unknown nature are also grouped into one class (No-ID).

Examining the table, one can note some important points: 
\begin{enumerate}
\item The catalog contains 4618 identified sources and 2577 non--identiffied sources
 (that is fractions of 64\% and 36\% of the total).

\item Most of the identified sources (4028 out 4618) are extragalactic
 objects, and almost all of them (4006 out of 4028)
 are Active Galactic Nuclei. The other extragalactic objects are
 6 normal galaxies, 8 starburst galaxies and 6 Seyfert galaxies.

\item A total of 590 sources are identified as Galactic objects,
 that belong to several different classes. 

\item A study of the latitude distribution of the 2577 sources without
 identification allows to estimate that approximately $1500\pm 100$
 of them are Galactic and $1100 \pm 100$ are extragalactic.

\item Combining the numbers of identified Galactic sources with the estimate
 of the number of resolved but non--identified Galactic sources
 one can conclude that the 4FGL catalog contains 
 approximately 2100 Galactic sources.

\item The cutoff functional form is used for the fits
 of 276 sources, that is only 3.8\% of the total.
 Almost all the sources fitted with this functional form
(with only 7 exceptions) are pulsars (PSR) and millisecond
pulsars (MSP). Conversely, almost all sources classified as PSR or MSP are
fitted with the cutoff form.

\item A large number of sources (42.8\%) are fitted with
 the log--parabola form, that is with a curved,
 gradually softening spectral shape.
\end{enumerate}

To study the global properties of the ensemble of
all Galactic sources it is necessary to take into account
for the fact that, as discussed above, only
28\% of the total (590 out of a total of approximately 2100)
are identified and associated to an astrophysical class.
In the following, to study the properties of the Galactic
sources including also the unidentified ones,
we will consider a set of 917 ``disk selected Galactic sources'',
that is all sources in the latitudes interval 
$|b| < 3^\circ$ that are not identified as extragalactic.
In this region the 4FGL catalog contains 1007 sources, 
of which 90 (all in the AGN class) are identified as extragalactic
and 289 as Galactic. The remaining set of 628 sources contains only
a small number of extragalactic objects of order $18\pm 2$.
This estimate has been obtained noting that
in the Galactic polar regions ($|\sin b| > 0.75$,
where the number of Galactic sources is very small),
81\% of the source are identified as extragalactic and 19\%
remain without identification, and assuming that the ``efficiency''
to identify an extragalactic source is approximately equal in the disk and polar regions.

One should keep in mind the fact that the small $|b|$ selection discussed
above includes less than half of the Galactic sources in the catalog,
and due to their spatial distribution, sources observed at large
and small latitudes are at different average distances and could
in principle have different properties.
However, our studies of the properties of identified Galactic sources
at small and large $|b|$ (see below) are consistent with the absence of
large differences. This issue should be further explored in future studies.

\subsection{Cutoff spectra}
It is striking that in the 4FGL catalog, with only seven exceptions,
the cutoff form is used exclusively for the fits of sources identified
as pulsars (PSR) and millisecond pulsars (MSP).
One of the exceptions is the Small Magellanic Cloud,
and in this case a large fraction of the gamma--ray emission
is probably produced by an ensemble of unresolved pulsars,
the other six exceptions are bright AGNs, and
in these cases the observed cutoffs are most likely
produced by absorption effects on gamma--rays traveling in extragalactic space.
Conversely, most of the sources in the pulsar and millisecond pulsar classes
(fractions of 94\% and 77\%, respectively) are fitted with the cutoff form,
and it is in fact natural to speculate that all
sources in the PSR and MSP classes have spectra with a
quasi--exponential cutoff that remains unobserved for the fainter objects.

\begin{figure}
\begin{center}
 \includegraphics[width=14.0cm]{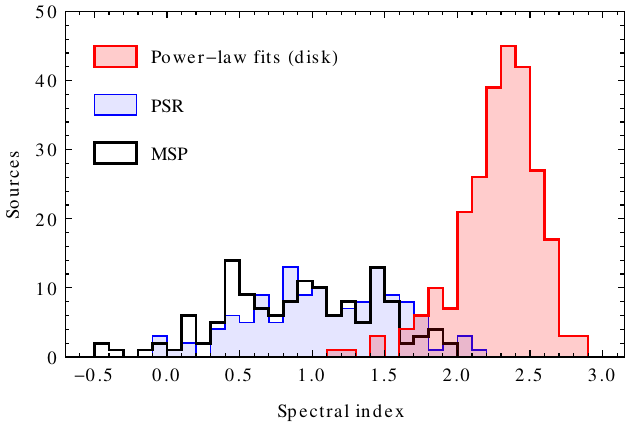}
\end{center}
\caption {\footnotesize
 The three histograms show the slope distribution for three sets of sources in the 4FGL catalog:
 (i) all sources in the disk region ($|b| < 3^\circ$, excluding extragalactic objects) that 
 that are fitted with the power--law form,
 (ii) all sources classified as pulsars (PSR) and fitted with the cutoff form,
 (iii) all sources classified as millisecond pulsars (MSP)
 and fitted with the cutoff form.
 For sources in the sets (ii) and (iii) the quantity in the histogram
 is low energy spectral index $\alpha_{\rm low}$ [see Eq.~(\ref{eq:cutoff1})].
\label{fig:index_pl_cutoff} }
\end{figure}

\begin{figure}
\begin{center}
\includegraphics[width=12.9cm]{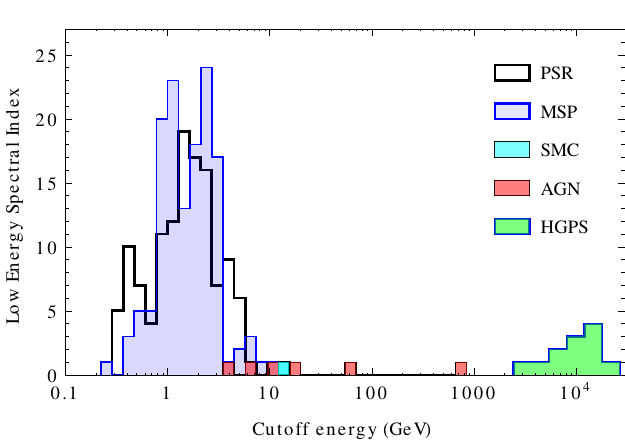}
\end{center}
\caption {\footnotesize
 Distribution of cutoff energy $E_{\rm cut}$ for the sources in the 4FGL--DR4 and HGPS catalogs
 that are fitted with the cutoff form.
 The different histograms show the distributions for different classes of objects.
\label{fig:cutoff_ecrit} }
\end{figure}

\begin{figure}
\begin{center}
\includegraphics[width=14.0cm]{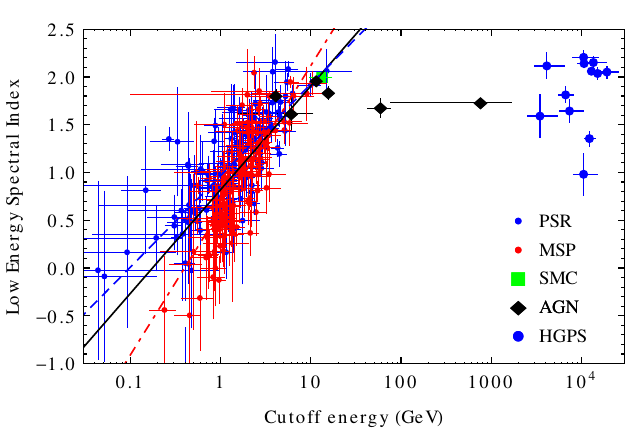}
\end{center}
\caption {\footnotesize
Scatter plot of the pair of fit parameters \{$E_{\rm cut}$,$\alpha_{\rm low}$\}
for all sources fitted with the cutoff form of Eq.(\ref{eq:cutoff1})
in the Fermi--LAT 4FGL--DR4 and HESS HGPS catalogs.
The different symbols show 
pulsars (PSR), millisecond--pulsars (MSP), the Small Magellanic Cloud (SMC),
and AGN measured by Fermi-LAT, and the sources in HGPS survey.
The lines show the best fit linear correlations between $\alpha_{\rm low}$ and $\log E_{\rm cut}$
for PSR+MSP (solid black curve), PSR (dashed blue), MSP (dot--dashed red).
The projections of this plot
%that is the distributions
%of $\log E_{\rm cut}$ and $\alpha_{\rm low}$ for this sets of sources,
are shown in Fig.~\ref{fig:cutoff_ecrit} and~\ref{fig:index_pl_cutoff}.
\label{fig:cutoff_distribution1} }
\end{figure}

More information about the cutoff spectra is given
in figures~\ref{fig:index_pl_cutoff}--\ref{fig:cutoff_distribution1}.
Fig.~\ref{fig:index_pl_cutoff} shows the distribution of the
low energy spectral index $\alpha_{\rm low}$, together with the
distribution of spectral index for the Galactic sources that are
fitted with a simple power--law form.
The distributions of $\alpha_{\rm low}$ for PSR and MSP sources
are very similar and show that the spectra are
very hard, with average index $\langle \alpha_{\rm low} \rangle \simeq 0.92$
for PSR and 1.08 for MSP, while 
the Galactic sources fitted with a simple power--law
form have much softer spectra with average index
$\langle \alpha \rangle \simeq 2.28$.

Fig.~\ref{fig:cutoff_ecrit} shows the distribution of the cutoff energy.
Again on finds that the distributions for PSR and MSP are
very similar, with $E_{\rm cut}$ taking values in the interval
0.3--3~GeV, with averages $\langle E_{\rm cut} \rangle \simeq 1.91$~GeV
for PSR and 1.95~GeV MSP's.

Fig.~\ref{fig:cutoff_distribution1} is a scatter plot of the
pair of fit parameters $\{\ln E_{\rm cut}, \alpha_{\rm low} \}$ for all sources
in the 4FGL catalog fitted with the cutoff form
(with different symbols indicating the different classes).
It is interesting to note that there is a significant linear correlation
between the fit parameters, with sources with larger $E_{\rm cut}$
having a harder spectrum below the cutoff.
The correlation coefficients, and the
best fit linear relationship between the parameters are given
in Table~\ref{tab:table_pulsars}.

\begin{table}[hbt]
\caption{\footnotesize
 Average values of the fit parameters for the Pulsar (PSR)
 and Millisecond Pulsars (MSP) sources
 in the 4FGL catalog that are fitted with the cutoff form of Eq.(\ref{eq:cutoff1}).
 The last two columns give the correlation coefficient, and the best fit for a linear
 relation between $\ln E^*$ and $\alpha_{\rm low}$.
\label{tab:table_pulsars}}
\renewcommand{\arraystretch}{1.29}

\vspace{0.25 cm}
 \begin{tabular}{|c || c | c | c || c | c |}
 \hline
 Class & Sources & $\langle E_{\rm cut} \rangle$ (GeV) & $\langle \alpha_{\rm low} \rangle $
 & Corr. Coef. & Correlation $\langle \alpha_{\rm low} \rangle (E_{\rm cut} ) $ \\
 \hline
PSR & 132 & $ 1.91 \pm 0.16 $ & $ 1.10 \pm 0.05 $ & 0.57 & $ 0.97 + 0.42 ~\log[E_{\rm cut}/{\rm GeV}] $ \\
MSP & 137 & $ 1.95 \pm 0.12 $ & $ 0.92 \pm 0.05 $ & 0.62 & $ 0.61 + 0.66 ~\log[E_{\rm cut}/{\rm GeV}] $ \\
PSR + MSP & 269 & $ 1.93 \pm 0.10 $ & $ 1.00 \pm 0.03 $ & 0.52 & $ 0.82 + 0.47 ~\log[E_{\rm cut}/{\rm GeV}] $ \\
 \hline
\end{tabular}
\end{table}
%%%%%%%%%%%%%%%%%%%%%%%%%%%%%%%%%%%%%%%%%%%%%%%%%%%%%%%%%%%%%%%%%%%%%%%%%%%%%%%%%%%%%

The cutoff functional form of Eq.~(\ref{eq:cutoff1})
contains the additional parameter $b$, that in most cases
(83\% for PSR and 96\% for MSP) has the fixed value $b = 2/3$,
and in all cases it is less than unity, so that
all cutoff fits are sub--exponential.

%The figure also shows the best fit parameters for the
%12 sources in the HGPS catalog that are fitted with an exponential cutoff form
%(more discussion in the next section).

The results shown in
figures~\ref{fig:index_pl_cutoff}--\ref{fig:cutoff_distribution1}, 
indicate that PSR and MSP sources have spectra with
a characteristic shape that reflects a characteristic emission mechanism.
It should however be noted that while the PSR and MSP emission
spectra have similar shape, they have 
very different absolute normalizations.
In fact the MSP sources have fluxes that are about 15 times smaller
and are also located at smaller distances.
Information about the average distance of PSR and MSP
sources can be infered from their latitude distributions.
Both distributions are centered on the Galactic equator
($b = 0$), but have very different widths,
that differ by a factor of nearly five
($\sigma_b \simeq 6.24 \pm 0.09$~degrees for PSR
and $29.63 \pm 0.14$~degrees for MSP), and the
disk region ($|b| < 3^\circ$) contains 7.3\% of the PSR
and 74\% of the MSP
(note that for an isotropic distribution the width is
$\sigma_b = 39.17^\circ$ and the disk region contains 5.2\% of the sources).
These results indicate that the observed millisecond pulsars are
distributed nearly isotropically in the sky
and therefore are very close to the Solar System,
and much fainter and more numerous than ordinary pulsars.

\subsection{Power--law and log--parabola spectra}
Most of the Galactic sources are fitted with the power--law
and log--parabola forms, with best fit parameters that take values
in wide intervals. The spectral index distribution of the power--law fits is shown in 
Fig.~\ref{fig:index_pl_cutoff} with values in the range from 1.1 to 2.8
and average $\langle \alpha \rangle \simeq 2.28$.
As already noted these spectra are significantly softer than those of
PSR and MSP sources below the cutoff energy.

\begin{figure}
\begin{center}
\includegraphics[width=14.0cm]{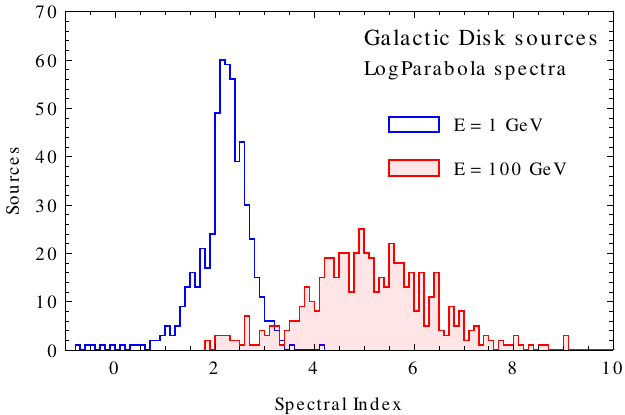}
\end{center}
\caption {\footnotesize
Spectral index distributiuons at $E =1$~GeV and $E = 100$~GeV
for the sources in the Fermi--LAT 4FGL--DR4 catalog that are
in the disk region ($b< 3^\circ$) and are fitted with the log--parabola form
(excluding sources associated to extragalactic objects). 
\label{fig:galactic_lp_index} }
\end{figure}

The spectral index for fits of the logparabola is proportional to $\log E$
[see Eq.~(\ref{eq:alpha_log--parabola1})]. For all sources
in the 4FGL catalog the curvature is positive so that
the spectra are softening and the spectral index increases with energy.
Fig.~\ref{fig:galactic_lp_index} shows the spectral index distributions
for all disk selected Galactic sources that are fitted with the log-parabola
at two values of the energy (1 and~100~GeV).
The average spectral index grows rapidly with energy 
taking values $\langle \alpha (E)\rangle = 2.18$, 3.64 and 5.10 for $E=1$, 10 and 100~GeV.
The average increase of the spectral index in a decade of energy
is $\langle \Delta \alpha \rangle = 2 \langle \beta \rangle \, \ln 10 \simeq 1.46$.
(with $\langle \beta \rangle$ the average curvature).

\begin{figure}
\begin{center}
\includegraphics[width=12.9cm]{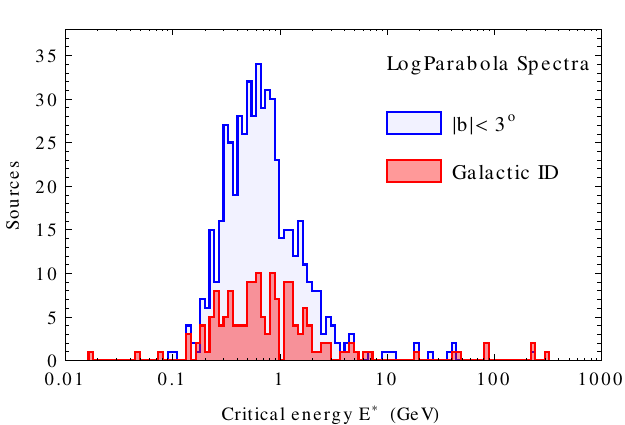}
\end{center}
\caption {\footnotesize
Distribution of $\log E^*$ (with $E^*$ the critical energy)
for gamma--ray sources in the Fermi 4FGL-DR4 that are fitted with the logparabola
form. The two histograms are for sources in the disk region
($b < 3^\circ$), excluding objects that are identified as extragalactic,
and for all objects that are identified as Galactic.
\label{fig:gal_lp_ecrit} }
\end{figure}

The shape of a log--parabola flux 
is described by two parameter: the critical energy $E^*$ and the curvature $\beta$.
%The distributions of these parameters for Galactic sources
%are shown in Figs.~\ref{fig:gal_lp_ecrit}--\ref{fig:galactic_lp_corr}. 
The critical energy distributions for the
Galactic sources fitted with the log--parabola form
is shown in Fig.~\ref{fig:gal_lp_ecrit}.
It is remarkable that $E^*$ can take values in a
very wide interval that spans three orders of magnitude
from less than 0.1~GeV to more than 100~GeV.

\begin{figure}
\begin{center}
\includegraphics[width=12.9cm]{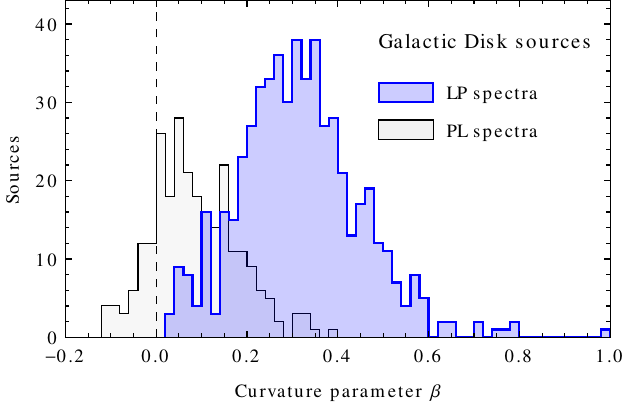}
\end{center}
\caption {\footnotesize
Distribution of the curvature parameter $\beta$
for disk--selected Galactic sources ($b < 3^\circ$).
The histogram with darker shading is for sources that have best fits
of the log--parabola form. The histogram with lighter shading gives
the curvature parameter for sources that are classified as having
power--law spectra.
\label{fig:beta_gal} }
\end{figure}

The distribution of the curvature parameter
for disk selected Galactic sources is shown in 
Fig.~\ref{fig:beta_gal} and has average $\langle \beta\rangle \simeq 0.317$ and 
width $\sigma_\beta \simeq 0.13$.
The 4FGL catalog gives a log--parabola fit also for sources
which are listed as having best fits of power--law form,
and Fig.~\ref{fig:beta_gal} also shows
the curvature distribution for these sources.
In 84\% of the cases the curvature is positive,
and the average is $\langle \beta\rangle \simeq +0.086$.
These results suggest that sources with best
fits of power--law and log--parabola form do not belong to separate classes
of astrophysical objects, and that the sources with PL best fits
form the low--curvature tail of a broad curvature distribution.

\begin{figure}
\begin{center}
\includegraphics[width=14.0cm]{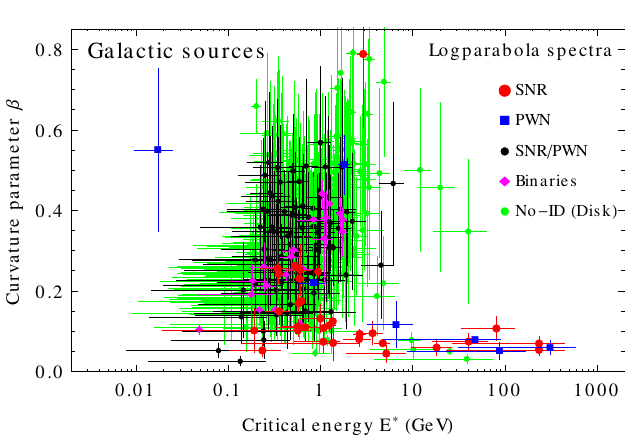}
\end{center}

\caption {\footnotesize
 Scatter plot of the fit parameters $\{E^*, \beta\}$ for the Galactic gamma--ray sources
 in the Fermi--LAT 4FGL-DR4 catalog that are fitted with the log--parabola form.
 All sources in the catalog associated to Galactic objects are included
 in the figure, with different symbols indicating sources in different
 classes [Supernova remnants (SNR),
 Pulsar Wind Nebulae (PWN), SNR/PWN sources, binary systems]. In addition also
 unidentified sources in the disk region $|b| < 3^\circ$ are included.
\label{fig:galactic_lp_corr} }
\end{figure}

Fig.~\ref{fig:galactic_lp_corr} is a scatter plot of the pair of parameters
$\{\log E^*, \beta\}$ for different classes of objects fitted with the log--parabola form.
It can also be noted that there is an intriguing hint of
a bimodal distribution, with one component formed by sources
with large curvature ($\langle \beta \rangle \approx 0.35$) and
small critical energy ($\langle E^* \rangle \sim 1$~GeV),
and a second component, formed mosty by SNR and PWN sources,
with smaller curvature and larger $E^*$.

\subsection{Cumulative Galactic source spectrum}
As a first step toward the construction of the
Milky Way source spectrum we have calculated the sum
of the fits to Galactic sources in the 4FGL catalog.

\begin{figure}
\begin{center}
\includegraphics[width=14.0cm]{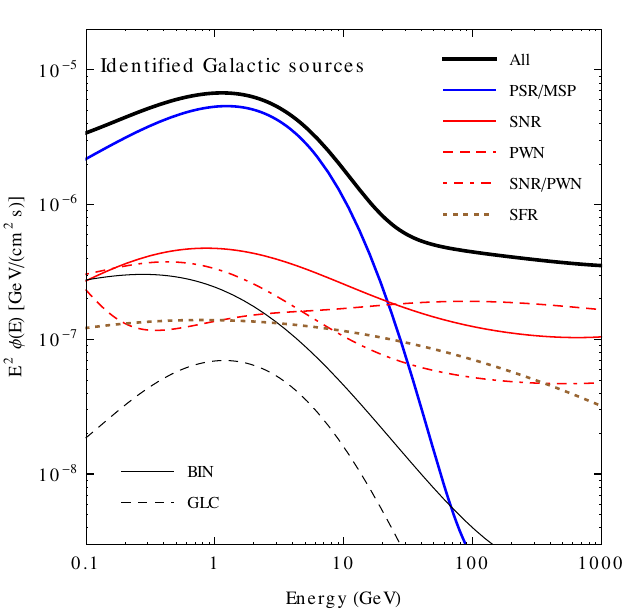}
\end{center}
\caption {\footnotesize
Cumulative spectrum of all sources in the 4FGL--DR4 catalog that are
identified as Galactic. The thick black line shows the sum for all sources.
The other lines show the sum over sources that have different classifications:
pulsars and millisecond pulsars (PSR/MSP), Supernova remnants (SNR),
Pulsar Wind Nebulae (PWN), mixed sources SNR/PWN, star formation regions (SFR),
Binary star systems and Galactic clusters.
\label{fig:galactic_id_spectra} }
\end{figure}

Fig.~\ref{fig:galactic_id_spectra} shows the spectrum obtained
summing the best fits to all sources in the 4FGL catalog that
are identified as Galactic (excluding novae and the Galactic Center).
The figure also shows the cumulative spectra obtained summing the
fits of sources that belong to seven different classifications: 
pulsars and millisecond pulsars (PSR + MSP),
supernova remnants (SNR), pulsar wind nebulae (PWN),
composite sources (associated with the overlap of a SNR and a PWN,
and labeled SPP in the catalog),
sources associated to star forming regions (SFR),
binary star systems (BIN), and globular star clusters (GLC).

\begin{figure}
\begin{center}
\includegraphics[width=14.0cm]{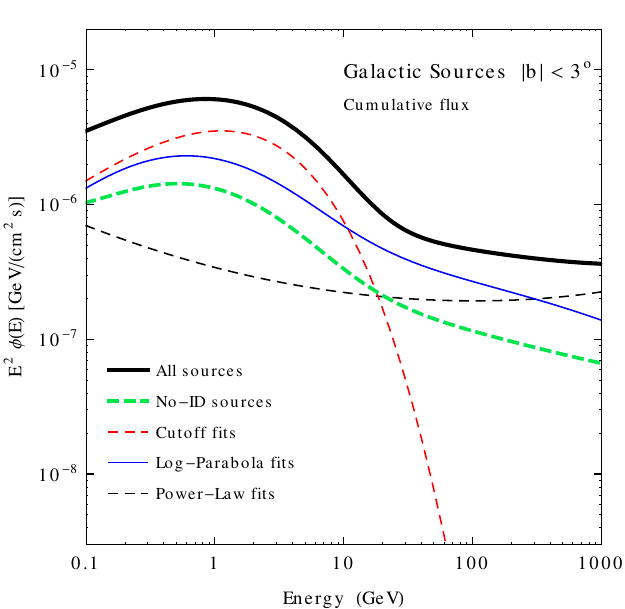}
\end{center}
\caption {\footnotesize
Cumulative spectrum of the gamma--ray sources in the 4FGL--DR4 catalog
that are in the disk region ($|b| < 3^\circ$), excluding objects identified
as extragalactic.
The thick, black, solid line shows the sum for all sources, 
the thick, green, dashed line the contribution
of sources that are not identified, the other curves show the contributions
of sources fitted with the three functional forms (cutoff, log--parabola and
power--law).
\label{fig:disk_spectra} }
\end{figure}

Fig.~\ref{fig:disk_spectra} shows the cumulative spectra 
of the 917 disk selected Galactic sources and of the
subset of 628 sources that have no identification.
The figure also shows the partial sums obtained summing
the fits to the sources fitted with the 
three functional forms used in the catalog
(power--law, log--parabola and cutoff).

Examining the spectra in Figs.~\ref{fig:galactic_id_spectra}
and~\ref{fig:disk_spectra}, one can make the following remarks:
\begin{enumerate}
\item The contributions of pulsars (PSR) and millisecond pulsars (MPS) dominate
the cumulative spectrum for $E \lesssim 20$~GeV. As discussed above,
% and illustrated
%in Figs. ~\ref{fig:index_pl_cutoff}--\ref{fig:cutoff_distribution1}
these sources have a characteristic spectral shape
(a hard power--law at low energy and a sub--exponential cutoff
with $E_{\rm cut}$ lower than few GeV) and the sum of these spectra
results in a flux of similar shape: very hard at low energy 
and with a sharp cutoff for $E_{\rm cut} \sim 3$~GeV. 

\item The cumulative spectrum of the Galactic sources that are not pulsars
overtakes the pulsar contribution at $E \simeq 20$~GeV,
and has a smooth shape, with a slope that in the entire energy range 0.1--$10^3$~GeV
remains approximately constant in a narrow range $\alpha (E) \approx 2.$--2.3.

\item
Four classes of sources: supernova remnants (SNR),
pulsar wind nebulae (PWN), composite sources (SNR/PWN)
and sources associated to star forming regions (SFR) give the largest
contributions the cumulative flux (excluding pulsars)
with spectra of the same order of magnitude and of roughly similar shape.

\item Smaller and softer contributions are given by sources
 classified as binary star systems (BIN) and globular star clusters (GLC).

\item Non--identified sources account for approximately 70\% of the number of
resolved Galactic sources, but they are on average fainter than the identified ones,
and account for only approximately 20--30\% of the cumulative flux.
Their spectral shape is roughly consistent with what can be obtained by a
combination of objects that belong to the same classes of the identified sources.

\item A comparison of the spectra of identified sources at large and small
 Galactic latitude does not reveal very large differences.
\end{enumerate}

\subsection{The formation of the Milky Way source spectrum}
The study of the cumulative spectrum of the Galactic sources
reveals that, excluding pulsars, one obtains a spectrum that has
a very smooth shape, with a spectral index that remains approximately constant
over a wide energy range, extending for nearly four orders of magnitude
(0.1--10$^3$~GeV). This smooth shape emerges from the
combination of spectra of individual sources that have a variety of different shapes.

\begin{figure}
\begin{center}
\includegraphics[width=12.5cm]{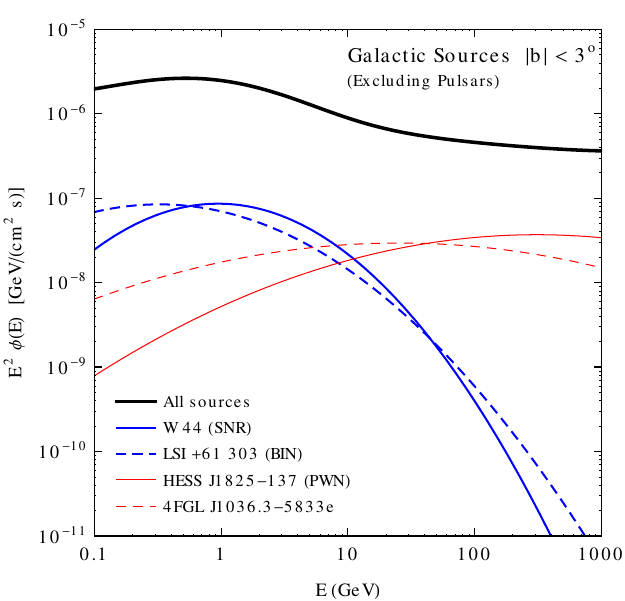}
\end{center}
\caption {\footnotesize
The thick line shows the cumulative flux obtained by summing the fits of all 
sources in the 4FGL catalog that are in the sky region $|b| < 3^\circ$
(excluding pulsars and extragalactic objects).
The other lines show the spectra of four sources individual sources
chosen as the objects that have the highest fluxes at $E=1$~GeV and $E = 100$~GeV
(excluding the Cygnus Cocoon).
\label{fig:gal_example} }
\end{figure}

The fact that the Galactic gamma--ray sources have different
spectral shapes implies that the set sources that give the largest contributions
to the cumulative flux, is continuously changing with energy.
An illustration of this concept can be seen
in Fig.~\ref{fig:gal_example}.
In this figure the thick line shows the cumulative spectrum obtained
summing the fits to the sources in the sky region $|b|< 3^\circ$,
excluding extragalactic objects, pulsars, novae and the Galactic center,
and the other four lines show the fits of four
individual sources, that have been chosen as the two objects that
(after the Cygnus Cocoon, a very bright object that is likely to be composite)
have the highest flux at $E = 1$ and 100~GeV.
All these four sources are fitted with the log--parabola form.

One can note that the two brightest sources at $E=1$~GeV
(the supernova SNR W~44 and the microquasar LSI+61~303)
have a rather small flux at 100~GeV,
and conversely the two brighest sources at 100~GeV
(the pulsar wind nebula HESS~J1825-137, and an unidentified object)
are faint at 1~GeV.
One can also note the cumulative spectrum obtained
summing the contributions of all Galactic sources in the sky region
considered is significantly smoother that the spectra of the individual
sources, in the sense that its spectral index remains
approximately constant in a wide energy interval, while
the spectral indices of the individual sources undergo
much larger variations.

It is interesting to note that if cumulative spectrum can be approximated
by a simple power law with a constant spectral index $\alpha_{\rm cum}$,
the relative contribution [that is the ratio $\phi_s(E)/\phi_{\rm cum} (E)$]
of a source $s$ with a specxtrum of log---parabola form
is energy dependent, increasing at low $E$ (when the source spectrum is harder),
and decreasing at high $E$ (when the source spectrum is softer).
The relative contribution of the source takes its maximum value
at the energy $E^\dagger$ that satisfies the condition that the slopes
of the source an cumulative spectra are equal:
$\alpha_s(E^\dagger) = \alpha_{\rm cum}$.
If the cumulative spectrum has slope $\alpha_{\rm cum} \approx 2$,
as it is in fact the case in the energy range we are discussing,
then the energy of the maximum contribution of a source of log--parabola form
is $E^\dagger \simeq E^*$ with $E^*$ the source critical energy.
One therefore expects that the sources that dominate the
cumulative spectrum at energy $E$ are those with critical
energy $E^* \approx E$.
One can in fact note that the two brightest sources in the disk
at $E \approx 1$~GeV have critical energies ($E^* =0.94$ and 0.32~GeV)
of the same order, while the two brighest sources at 100~GeV
have much higher critical energies ($E^* = 309$ and 25~GeV).

The discussion developed above suggests that the total Galactic source flux
is formed in a non trivial way by the contributions of individual objects
that have spectral shapes that are different from each other,
and different from the shape of the total flux.
This implies that the set of individual objects that give the largest 
contributions to the total source evolve with energy.

\begin{figure}
\begin{center}
~~~\includegraphics[width=12.9cm]{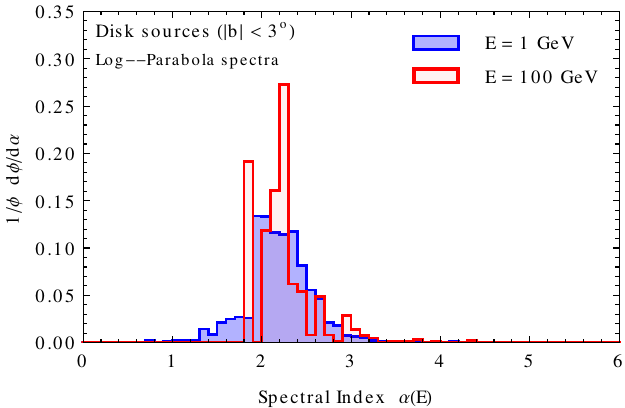}
\end{center}
\caption {\footnotesize
Weighted spectral index distributions 
at energies $E =1$~GeV and $E = 100$~GeV of the sources
in the Fermi--LAT 4FGL--DR4 catalog that are in the disk region ($b< 3^\circ$)
and are fitted with the log--parabola form
(objects identified as extragalactic and the Cygnus Cocoon have been excluded).
The weights are proportional to the flux of each source
at the selected energy.
\label{fig:gal_lp_w_index} }
\end{figure}

In the energy range of the Fermi observations,
excluding pulsars that have a characteristic spectral shape,
the total Galactic source spectrum has a nearly constant spectral index,
and most (or perhaps all) of the individual sources have spectra reasonably
well approximated by the log--parabola form with a broad range of shape
parameters. The contributions of these individual sources to the total has then
the following properties:
\begin{itemize}
\item [(a)] The sources that give the largest contributions to the total flux
 at energy $E$ are those that have spectral index
 $\alpha_s(E)$ approximately equal to the slope $\alpha_{\rm cum}$ of the total source flux.
\item[(b)] Since the total source flux has a slope $\alpha_{\rm cum} \approx 2$,
 the sources that dominate the total flux at energy $E$ are those
 with critical energy $E^* \approx E$.
\end{itemize}
It is possible to test the two statements above with the 4FGL data.
To test the statement (a) one can construct the distribution
$d\phi_{\rm cum}(\alpha;E)/d\alpha$ that describes
how the cumulative flux at energy $E$ is formed by the contributions
of sources with spectral index $\alpha_s(E) = \alpha$. This distribution
can be calculated as:
\begin{equation}
 \frac{d\phi_{\rm cum} (\alpha; E)}{d\alpha} = \sum_s \phi_s (E)
 ~\delta [\alpha - \alpha_s(E)] 
\label{eq:sumalpha}
\end{equation}
where the summation is over all the individual sources
(labeled by the subscript $s$).
Fig.\ref{fig:gal_lp_w_index} shows this distribution
%$d\phi_{\rm cum}/d\alpha(\alpha; E)$
for two values of the energy (1~and 100~GeV).
Examining the figure one can see how the distributions
for the two energies are very similar. In 
both cases only the sources with slopes $\alpha_s (E)$ in
a narrow range of values centered around $\alpha \simeq 2.3$
(that is the approximately constant slope of the cumulative spectrum)
contribute to the total flux.
It is instructive to compare these histograms,
where the contribution of each source
is weighted by its flux,
with the spectral index distributions shown in Fig.~\ref{fig:galactic_lp_index},
where all sources are weighted equally. In this case
the distributions for the two energies are very different 
(with averages 2.18 and 5.10 at 1 and 100~GeV respectively) and much broader.

\begin{figure}
\begin{center}
\includegraphics[width=12.9cm]{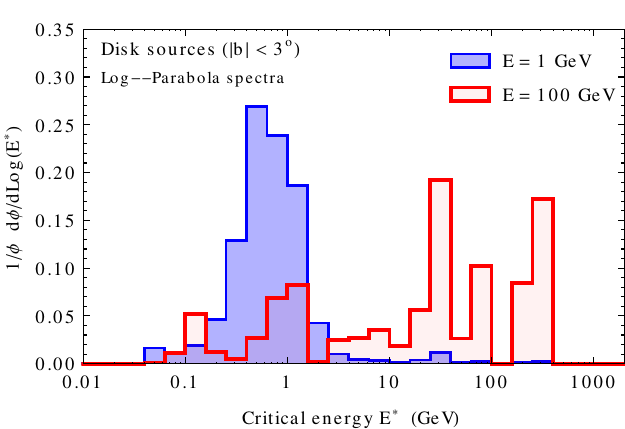}
\end{center}

\caption {\footnotesize
Distributions of $\log E^*$ (with $E^*$ the critical energy) 
for the gamma--ray sources in the Fermi 4FGL-DR4 that are in the disk region $b < 3^\circ$
(excluding objects identified as extragalactic and the Cygnus Cocoon)
and are fitted with the log--parabola form.
The two histograms are obtained weighting the 
contribution of each source with its flux at energy
$E=1$~GeV and $E = 100$~GeV.
\label{fig:gal_lp_w_ecrit} }
\end{figure}

To verify statement (b) one can construct the distribution
$d\phi_{\rm cum} (E^*; E)/d\log E^*$ that describes
how the cumulative flux at energy $E$ is formed by the contributions
of sources that have critical energy $E^*$:
\begin{equation}
 \frac{d\phi_{\rm cum} (E^*; E)}{d\log E^*} = \sum_s \phi_s (E)
 ~\delta [\log E^* - \log E^*_s] ~
\label{eq:sumestar}
\end{equation}
Fig.\ref{fig:gal_lp_w_ecrit} shows this distribution
% (\ref{eq:sumestar})
%$d\phi_{\rm cum}/d\log E^* (E^* ; E)$ 
for two values of the gamma--ray energy: $E=1$~GeV and $E=100$~GeV.
Examining the figure one can see that
at the lower energy the cumulative flux is mostly formed mostly by
sources with critical energy of order 1~GeV
(with weighted average $\langle E^* \rangle \simeq 2.4$~GeV).
At the higher energy ($E = 100$~GeV),
the cumulative spectrum is dominated by sources with much larger critical energy
(with weighted average $\langle E^* \rangle \simeq 67$~GeV).
Note that in this case a smaller number of sources make up
the cumulative spectrum. Twelve sources with critical energy $E^* > 10$~GeV
generate approximately 66\% of the total flux.

%These results indicate that it is necessary to clearly distinguish between
%the spectral shape of the ensemble of the Galactic gamma ray sources,
%and the spectral shapes (note the plural) of the individual sources.
%The commonly used assumption that gamma--ray sources that belong to the same class
%have spectra of approximately equal shape is seriously incorrect.

\section {High energy measurements of Galactic gamma--ray sources}
\label{sec:tev_sources}
Observations of gamma--rays at high energies ($E \gtrsim 1$~TeV) have been
recently made by ground--based telescopes, and in the
following we will discuss the results contained in the three 
catalogs introduced in section~\ref{sec:catalogs}:
HGPS, 3HAWC and 1LHAASO.
Comparing results obtained from studies of different catalogs
one should keep in mind that they cover different sky 
regions with different sensitivities.
These differences are obviously very important 
for the absolute normalization of the cumulative spectra
obtained by summing the fluxes of all the sources in a catalog,
but are likely to be much less significant for the shape of these spectra. 

\begin{figure}
\begin{center}
\includegraphics[height=6.3cm]{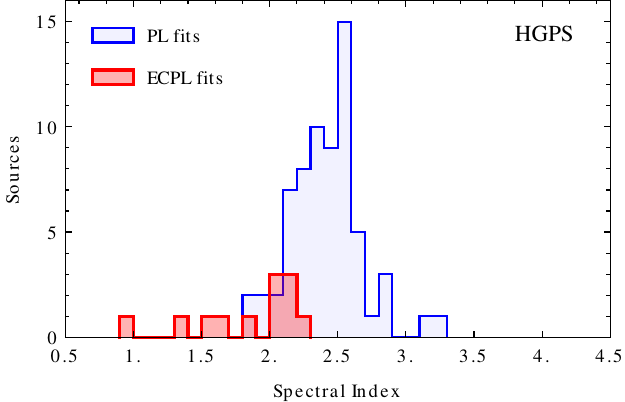}
\end{center}
\begin{center}
\includegraphics[height=6.3cm]{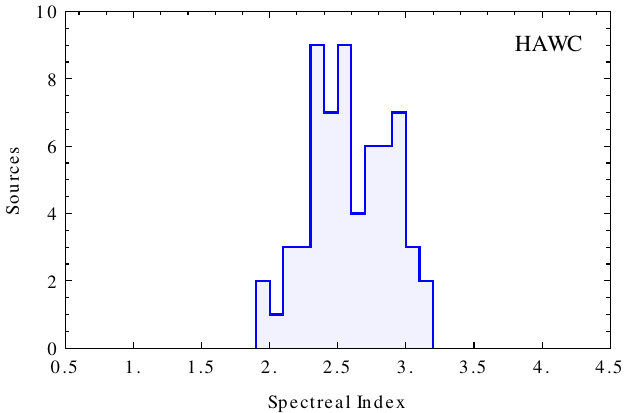}
\end{center}
\begin{center}
\includegraphics[height=6.3cm]{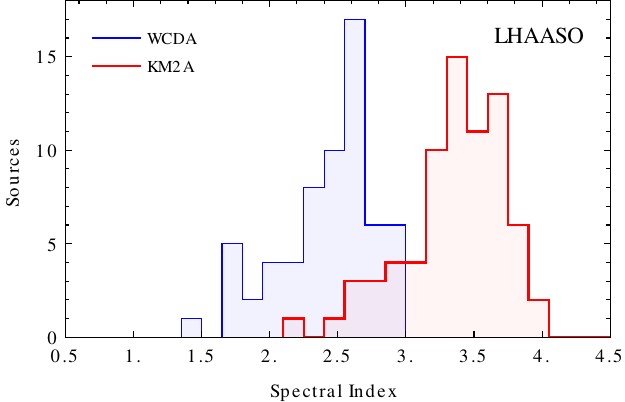}
\end{center}
\caption {\footnotesize
Spectral index distributions
of the gamma--ray sources in three high energy catalogs:
HGPS (top panel), HAWC 1st catalog (center panel) and LHAASO 1st catalog
(bottom panel).
In the top panel the distributions for the HGPS sources 
that are fitted with the power--law and cutoff forms are shown separately.
In the bottom panel the two histograms show the spectral index distributions of the
LHAASO sources observed by the WCDA and KM2A arrays.
\label{fig:catalogs_slopes} }
\end{figure}

The HAWC and LHAASO catalogs fit the sources with a power--law form,
while in the HGPS catalog the fit can be a
simple power--law (PL, for 66 objects),
or a power--law with an exponential cutoff (ECPL, for the remaining 12 objects).
The spectral index distributions for the high energy gamma--ray catalogs 
are shown in Fig.~\ref{fig:catalogs_slopes},
and the averages, widths and extensions of the distributions are listed
in Table~\ref{tab:table_slopes_high}.

Examining the table some important results emerge.
The first one is that the observations of the three telescopes in the 1--10~TeV
energy range are in reasonable good agreement,
and obtain similar averages of 2.33, 2.58, and 2.41 for HGPS, 3HAWC and LHAASO--WCDA,
respectively, a slope that is a larger (implying softer spectra)
than what is observed at lower energies in the Fermi--LAT catalog.
The sources observed by LHAASO--KM2A at higher energies
($E \ge 25$~TeV) have much softer spectra, 
with average slope $\langle \alpha \rangle \simeq 3.38$.
In all cases the spectral index distributions are broad,
indicating that the sources can have spectra of quite different shape.

One should also note that in the HESS catalog 12 out of 78 sources 
(a fraction 15.4\%) have been fitted with the ECPL form
(a power--law with an exponential cutoff).
The cutoff energy distribution is shown in Figs.~\ref{fig:cutoff_ecrit}
and~\ref{fig:cutoff_distribution1} and
extends from~3.5 to~19.2~TeV with average $\langle E_{\rm cut} \rangle \simeq 10.5$~TeV.
As discussed in section~\ref{sec:fermi}, in the Fermi--LAT catalog
only the spectra of pulsars and millisecond pulsars sources have sharp cutoffs,
while the spectra of other Galactic sources soften more gradually.
It is very important to establish how many sources exhibit
(quasi)--exponential cutoffs, and at which energies.
This problem can be studied by examining the LHAASO observations 
that thanks to its two arrays can study the spectral shape of gamma--ray sources
over a very wide energy range spanning about
three decades (from 1 to $10^3$~TeV).

\begin{figure}
\begin{center}
\includegraphics[width=8.50cm]{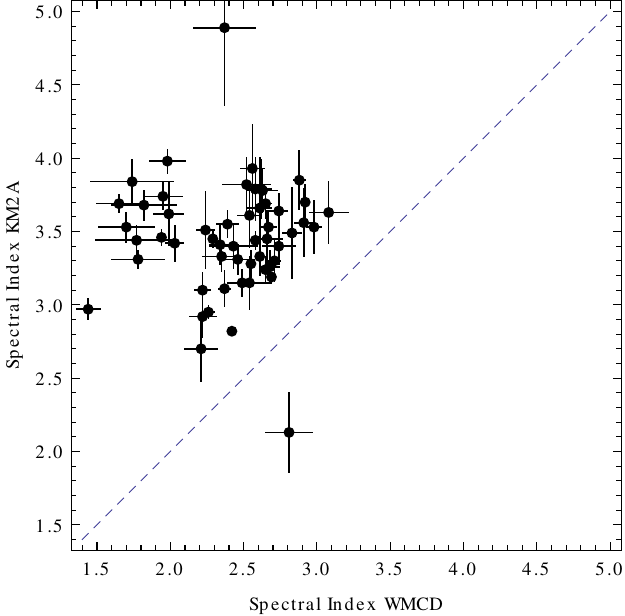}

\vspace{0.02 cm}
\includegraphics[width=8.50cm]{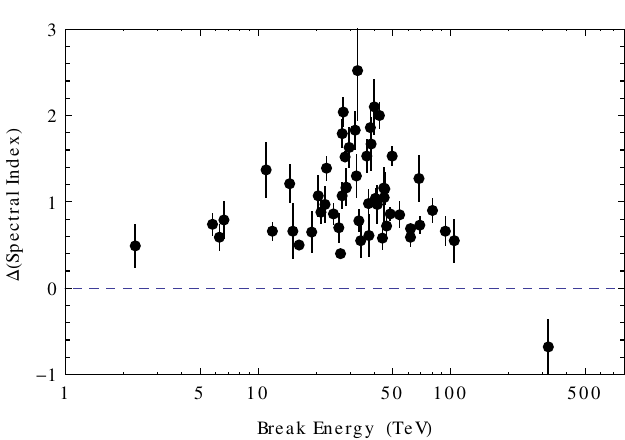}
\end{center}

\begin{center}
\includegraphics[width=7.0cm]{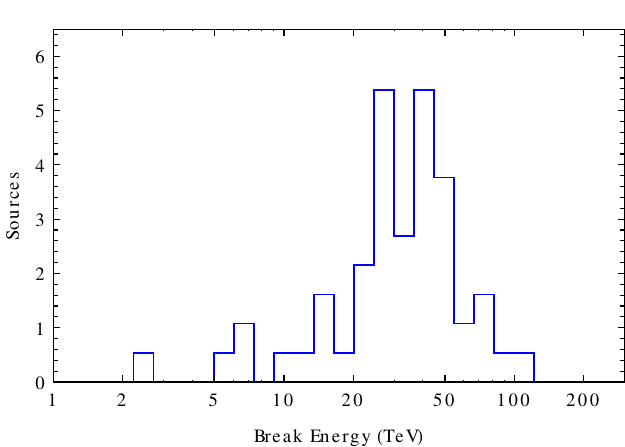}
~~~
\includegraphics[width=7.0cm]{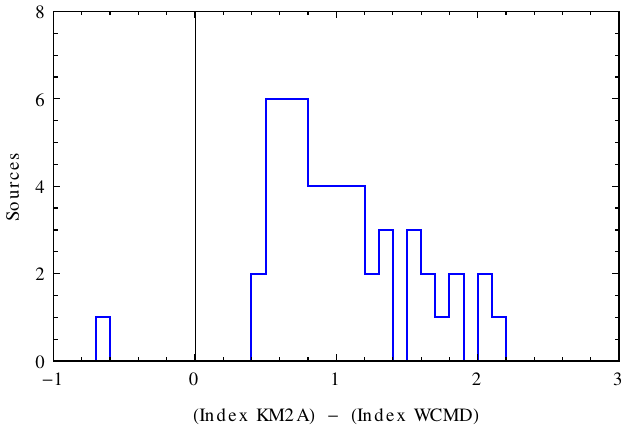}
\end{center}
\caption {\footnotesize In the upper panel the points with error bars show the
 spectral indices of the best fits to the observations by the WCDA and KM2A arrays
 of the LHAASO telescope for the 54 sources that are detected by both instruments.
 The dashed line correspons to equals indices in the two arrays, and
 it can be seen 
 that with only one exception, the measurements at higher energies by the
 KM2A array give a softer spectrum with a larger spectral index.
 In the central panel the points show, for the same 54 sources,
 the pair of quantities $\{E_b, \Delta \alpha\}$
 with $E_b$ the (unique) ``break energy'' where the best fits of the observations by 
 the two LHAASO arrays intersect each other,
 and $\Delta \alpha = \alpha_{\rm KM2A} - \alpha_{\rm WCDA}$ is the difference
 between the spectral indices in the two fits. In the bottom part of the figure
 the two panels show the distributions (always for the same 54 sources) of
 the break energy $E_b$ and
 of the difference in spectral index $\Delta \alpha$.
\label{fig:lhaaso_2detections}}
\end{figure}

In the first LHAASO catalog a total of 54 sources
are observed in both the WCDA and KM2A arrays. These spectra do not exhibit
a cutoff but they all, with one exception, have an important softening.
Figure~\ref{fig:lhaaso_2detections} shows (in the top panel)
a scatter plot of the pair of slopes $\{\alpha_W, \alpha_K\}$
obtained from fits to the WCDA and KM2A data for all these 54 sources.
With only one exception, the difference $\Delta \alpha = \alpha_K - \alpha_W$
is positive, indicating that the spectra undergo a softening. 
Two power--laws of different slope intersect at one ``break energy''.
The middle panel of Fig.~\ref{fig:lhaaso_2detections} shows
a scatter plot of $\Delta \alpha$ versus $E_{\rm break}$,
and the bottom panel the distributions of the two quantities.
The average break energy is 
$\langle \log E_{\rm break} \rangle \simeq \log [31~{\rm TeV}]$,
and for the slope difference distributions one finds 
average $\langle \Delta \alpha \rangle \simeq 1.04$
and width 0.54.

%%%%%%%%%%%%%%%%%%%%%%%%%%%%%%%%%%%%%%%%%%%%%%%%%%%%%%%%%%%%%%%%%%%%%%%%%%%%%%%%%
\begin{table}
 \caption{\footnotesize
 Parameters of the spectral index distributions for sources
 in the high energies gamma--ray catalogs
 (known extragalactic objects have been excluded).
\label{tab:table_slopes_high}}
\renewcommand{\arraystretch}{1.29}

\vspace{0.25 cm}
 \begin{tabular}{|l || c | c | c | c | c |}
 \hline
 Catalog & Sources & $\langle \alpha \rangle $ & $\sigma_{\alpha}$
 & $\alpha_{\rm min}$ & $\alpha_{\rm max}$ \\
\hline
HGPS (ECPL fits) ~ & 12 & $1.84 \pm 0.10$ & 0.36 & 0.92 & 2.20 \\
HGPS (PL fits) & 66 & $2.41 \pm 0.03$ & 0.26 & 1.80 & 3.27 \\
HGPS (all fits) & 78 & $2.32 \pm 0.04$ & 0.35 & 0.92 & 3.27 \\
HAWC & 63 & $2.58 \pm 0.04$ & 0.29 & 1.98 & 3.14 \\
LHAASO--WCDA & 64 & $2.45 \pm 0.05$ & 0.37 & 1.44 & 3.18 \\
LHAASO--KM2A & 75 & $3.38 \pm 0.05$ & 0.43 & 2.13 & 4.89 \\
 \hline
\end{tabular}
\end{table}
%%%%%%%%%%%%%%%%%%%%%%%%%%%%%%%%%%%%%%%%%%%%%%%%%%%%%%%%%%%%%%%%%%%%%%%%%%%%%%%%%

\begin{figure}
\begin{center}
\includegraphics[width=12.9cm]{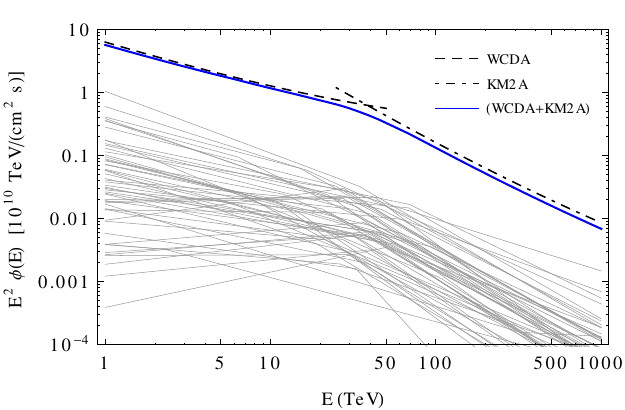}
\end{center}

\vspace{0.9 cm}
\begin{center}
\includegraphics[width=12.9cm]{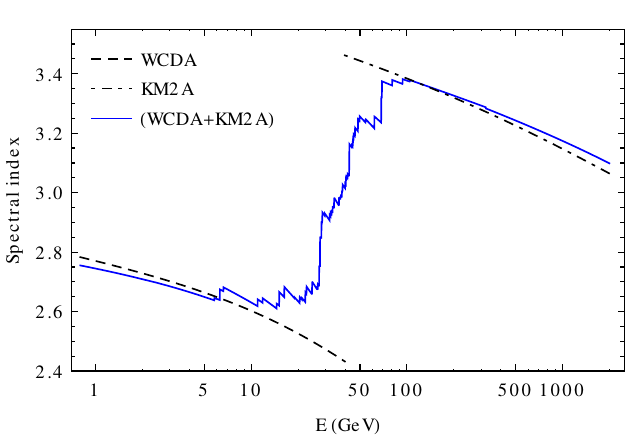}
\end{center}
\caption {\footnotesize
The top panel shows the cumulative spectra obtained by summing the fits to the sources
in the 1LHAASO catalog. The dashed (dot--dashed) line shows the
spectrum for the sources observed in the WCDA (KM2A) array
(excluding objects identified as extragalactic).
The thick solid line shows the cumulative spectrum for the 54 sources observed in both
arrays. In this case the fits to the observations of the two arrays have been
combined in the simple way described in the main text. 
The thin lines show the fits for the individual sources observed in both arrays.
The bottom panel shows the energy dependence of the index
of the cumulative spectrum shown above.
The dashed (dot--dashed) line is for sources observed in the
WCDA (KM2A) array, and the solid line is for sources observed by both arrays.
In this case the spectral index has 54 discontinuities at the break energies of the
fits to individual sources.
 \label{fig:lhaaso_2det_spectra}}
\end{figure}

For the sources that are observed by both the WCDA and KM2A arrays
it is obviously desirable to combine the low and high energy fits.
The simplest way is to simply connect the two power--laws of
the single--array fits having a discontinuous derivative
at the break energy.
The spectra obtained with this very simple algorithm
are shown in Fig.~\ref{fig:lhaaso_2det_spectra}.
Summing all these lines one obtains a cumulative spectrum 
that is shown as a thick line in Fig.~\ref{fig:lhaaso_2det_spectra}
and can be described, to a good approximation as the 
combination of two power--laws connected by a gradual ``knee--like'' softening
that develops in an energy interval of about half a decade,
from 30 to 100~TeV, with a step in spectral index of about one unit.
The figure also shows the cumulative spectra obtained by summing the fits of all sources
observed in the WCDA or KM2A array including those not observed in the other array.

The bottom panel in Fig.~\ref{fig:lhaaso_2det_spectra} presents the spectral index
of the cumulative spectra shown above.
Two lines show the slopes of the cumulative spectra
for the sources observed by the WCDA and KM2A arrays, and a
third line shows the slope of the spectrum that sums
the combined fits to the sources observed by both arrays.
In this last case the spectral index has a
discontinuity for all $E$ equal to the break energies
of the spectra included in the sum.

The observations of the LHAASO telescope
suggest that most of the TeV gamma--ray sources have spectra that extend
to very high energy, with shapes that deviates from a simple power--law form 
undergoing softenings in the 10--100~TeV energy range
that corresponds to large steps in spectral index
(with average $\langle \Delta \alpha \rangle \approx +1$).
The conclusion that most gamma ray sources
have a break or a softening in the 10--100~TeV energy range is
also consistent with the result that several sources 
are observed in one of the LHAASO arrays and not in the other.
In most of these cases the flux upper limit obtained by the array where the source
is not measured implies the existence of a spectral break, since
the non--detection implies that the spectrum must become
softer at higher energies (for sources observed only in WCDA),
or harder at lower energies (for sources only observed in KM2A).

\begin{figure}
\begin{center}
\includegraphics[height=5.6cm]{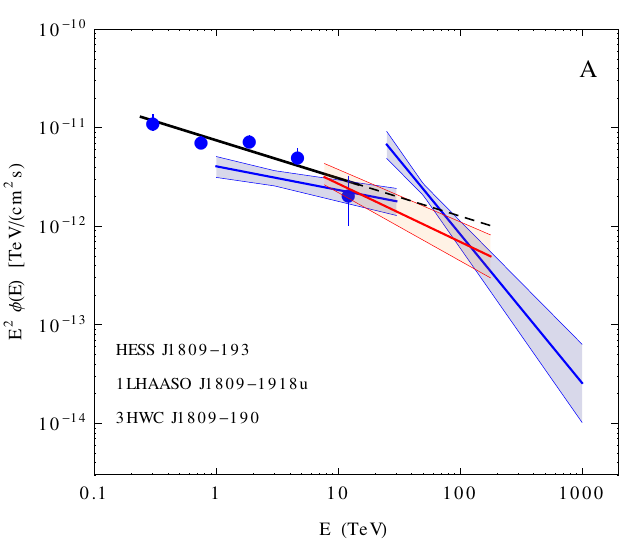}
~~~
\includegraphics[height=5.6cm]{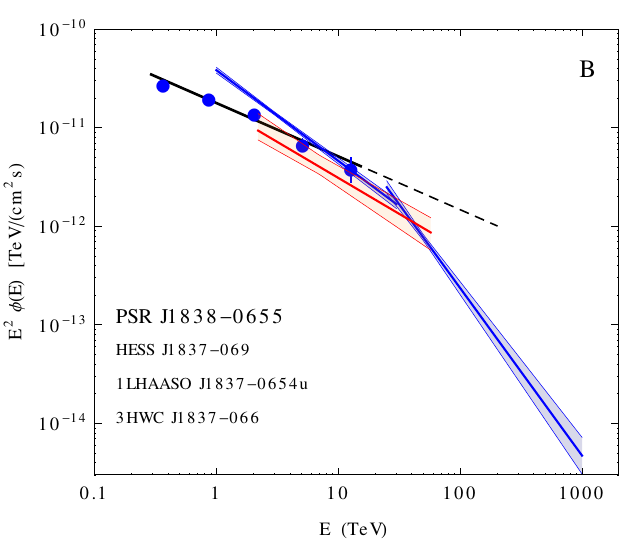}
\end{center}

\vspace{0.05 cm}
\begin{center}
\includegraphics[height=5.6cm]{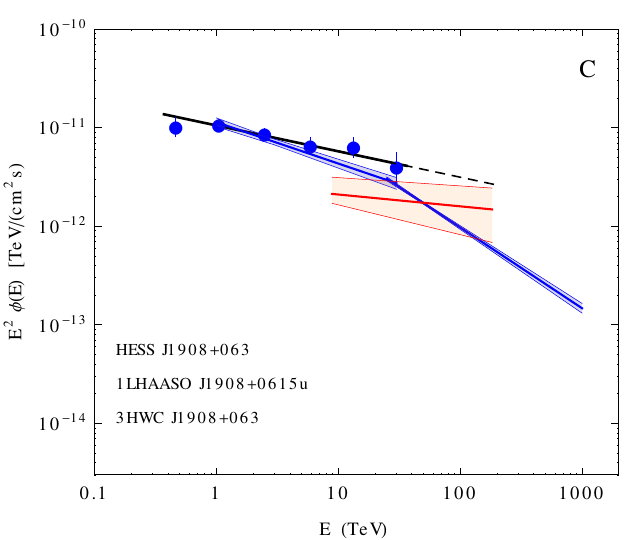}
~~~
\includegraphics[height=5.6cm]{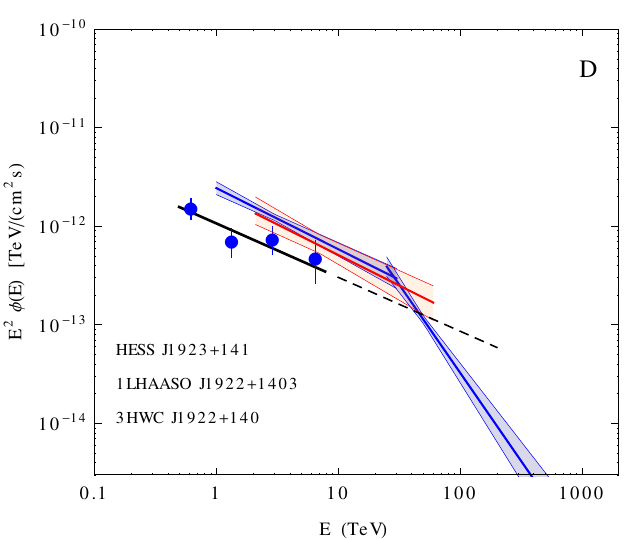}
\end{center}

\vspace{0.05 cm}
\begin{center}
\includegraphics[height=5.6cm]{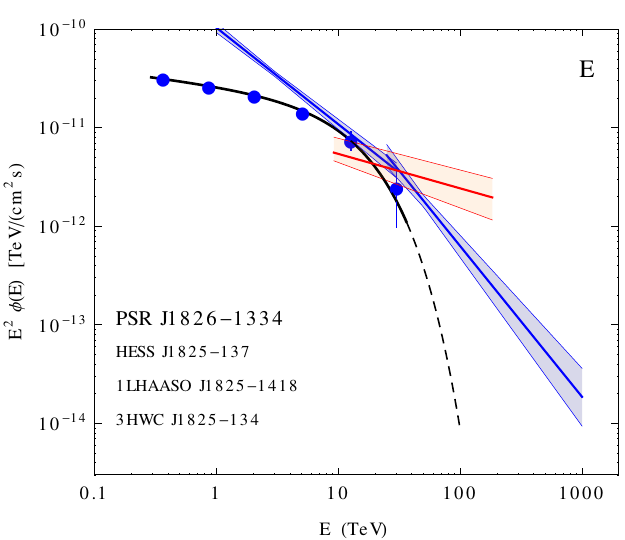}
~~~
\includegraphics[height=5.6cm]{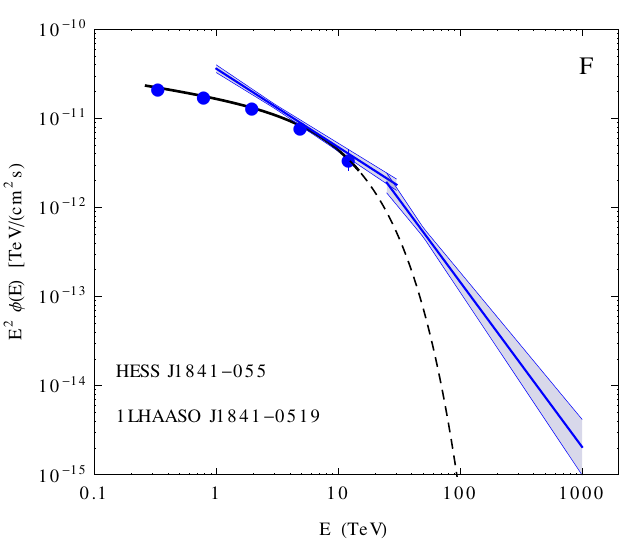}
\end{center}

\caption {\footnotesize
 Examples of the spectra of gamma--ray sources listed in the
 catalogs of the HESS, HAWC and LHAASO telescopes. The data of HESS (HGSP)
 is shown as points with error bars
 with the solid line showing the best fit, and the dashed line
 its extrapolation to higher energies. For the observations of the
 HAWC and LHAASO the figure shows, as shaded areas,
 the bests fits and their uncertainties.
 The HESS observations for the two sources shown in the bottom row of the figure
 have been fitted with the ECPL form (power--law with exponential cutoff), all other
 sources have been fitted with the simple power--law form.
\label{fig:source_spectra1} }
\end{figure}

The LHAASO observations of Galactic sources contained in the 1st catalog
do not find clear evidence for sharp, (quasi)--exponential cutoffs
in the spectra, and it is important to compare these results
with those of the HGPS catalog, where approximately 15\% of the sources
are fitted with the ECPL form.
Some of the HGPS sources are also in the LHAASO field of view,
and this allows to extend the flux measurements to higher energy
and study the presence of spectral softenings and of their structure
(a sharp (quasi)--exponential cutoff or a more gradual one).
Fig.~\ref{fig:source_spectra1} shows some examples of spectra
of gamma--ray sources observed by HESS, LHAASO and HAWC, where one can see
that the extension of the measurements to a wider energy interval
is important to determine the spectral shape of a source and establish
the existence of a cutoff.
A preliminary study suggests that some of the spectra fitted by HESS with the ECPL form,
when extended to higher energy, are better described by a more gradual softening,
but other do have sharp cutoffs. This important question requires additional studies.

\subsection{Matching the Fermi--LAT and high energy gamma--ray catalogs}
It is interesting to compare the catalog of gamma--ray sources
obtained by the Fermi--LAT telescope for $E \lesssim 1$~TeV 
with those obtained at higher energies by ground--based telescopes.
In making this comparison it is important to remember 
that ground--based telescopes observe only a limited region of the sky.

\begin{figure}
\begin{center}
\includegraphics[width=10.0cm]{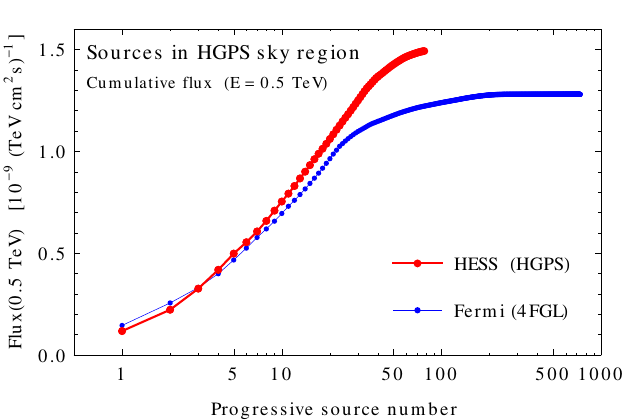}
\end{center}
\caption {\footnotesize
Cumulative fluxes at energy $E = 0.5$~TeV obtained by summing the fits
$\phi_j (E)$ (with $j \le N$) to the $N$ brightest sources at energy $E$ in
a catalog, and plotted as a function of $N$. The two sets of points (joined by a line)
are for the HGPS catalog (78 sources) and for the 4FGL catalog, including only
objects that are in the HGPS sky region and are not identified as
extragalactic (738 sources).
The total cumulative fluxes in the two cases differ by approximately 16\%. 
\label{fig:fermi_hess_cum}}
\end{figure}

\begin{figure}
\begin{center}
\includegraphics[width=10.5cm]{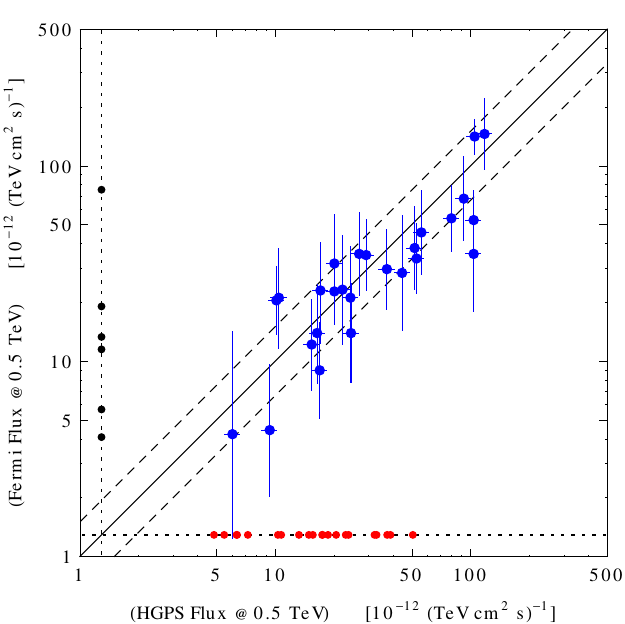}
\end{center}
\caption {\footnotesize
Matching between the brightest sources
at energy $E= 0.5$~TeV
[$\phi_j (E) \ge 4 \times 10^{-12}$~(cm$^2$~s~TeV)$^{-1}$]
in the HGPS catalog (32 objects) and in the 4FGL catalog
(including only objects that are in the HGPS sky region and are not identified
as extragalactic, 47 sources).
A match is obtained if the angular distance between the best fit positions
of two sources (in different catalogs) is less than 1.0 degree.
This algorithm yields 26 matches, that are shown as blue dots with error bars
in the $\{\phi_{\rm HESS}(E), \phi_{\rm Fermi} (E)\}$ plane
which gives the best fit fluxes at $E=0.5$~TeV for the two telescopes.
The non--matched sources (6 for Fermi and 21 for HESS)
are also shown as black and red dots, respectively.
\label{fig:fermi_hess}}
\end{figure}

In the following, in Figs.~\ref{fig:fermi_hess_cum} and~\ref{fig:fermi_hess}
we will compare the sources listed in the HGPS catalog
with the Fermi--LAT sources in the same sky region
\{$|b| < 3^\circ$, $ -110^\circ \le \ell \le +65^\circ$\}.
In this region, that is particularly important as it contains the Galactic center,
the HGPS catalog lists 78 sources, while the sources in the
Fermi--LAT catalog are one order of magnitude more numerous
(781 objects of which 43 are classified as extragalactic).
The cumulative spectra obtained by summing the fits
to all 78 HGPS sources and to the 738 sources in the 4FGL
catalog that are not classified as
extragalactic \footnote{This set of sources contains only a small
contamination, of order 7 to 10, of non--identified
extragalactic objects (see discussion in section~\ref{sec:fermi}. }
are very similar in shape and in absolute normalization.
The cumulative fluxes are shown in Fig.~\ref{fig:all_spectra}, where one can see
that the Fermi--LAT spectrum is only a little harder,
and the two spectra cross each other at $E \simeq 0.88$~TeV

This result, given the large difference in the number of resolved sources
in the same sky region might appear surprising,
but it can be undersood as the consequence of two facts: 
(1) the two telescopes have approximately the same sensitivity
to gamma--ray sources for $E \sim 1$~TeV, but different sensitivities
below and above this energy;
(2) the Galactic gamma--ray sources can have different spectral shapes, and
most of the sources resolved by Fermi--LAT are detected at lower energies,
and give negligible contributions to the cumulative flux in the TeV energy range.

To study this problem more quantitatively one can examine
Figs.~\ref{fig:fermi_hess_cum} and~\ref{fig:fermi_hess}.
To obtain Fig.~\ref{fig:fermi_hess_cum}
all sources in the HGPS sky region contained in the HGPS and 4FGL catalogs
have been sorted according to the differential flux
at the reference energy $E_0 = 0.5$~TeV (where both telescopes have a good sensitivity).
Then, for both catalogs, we have calculated
the cumulative fluxes $\phi_{{\rm cum},N}$ obtained by summing
the fluxes of the $N$ brightest sources. The
results are then been plotted as a function of $N$ in the figure.
Examining Fig.~\ref{fig:fermi_hess_cum} one can see that 
most of the Fermi--LAT sources give a negligible
contributions to the total flux at $E = 0.5$~TeV, so that the cumulative spectra
for the two catalogs at this energy are approximately equal,
differing by only 16\%.

Figure~\ref{fig:fermi_hess_cum} shows that a small number (of order 20)
of bright objects in both catalogs account for a large fraction of the
total source flux, and also suggests that many of these bright objects
are observed by both telescopes.
A detailed discussion of the matching between the Fermi--LAT and HESS catalogs
is beyond the scope of this paper, but we have performed a simple preliminary study
selecting the brightest sources in the two catalogs,
namely those with differential flux at the reference energy $E=0.5$~TeV
larger that the minumum value:
$\phi_{\rm min} = 4 \times 10^{-12}$~(cm$^2$~s~TeV)$^{-1}$.
This cut selects 32 Fermi--4FGL and 47 HGPS objects.
Sources in the two catalogs are then matched
if the angular distance between their sky positions is less than 1.0~degree.
This condition results in 26 matches,
leaving 6 sources in the 4FGL catalog and 21 in the HGPS catalog without a
match. The fluxes of all these sources are shown in Fig.~\ref{fig:fermi_hess}.

The result that sources that have the same flux at one reference energy
(in our discussion $E_0 =0.5$~TeV) can be resolved by one telescope and not by the
other and viceversa can be explained assuming that the gamma--ray sources
can have different spectral shapes.
The detection of a source by a telescope depends
on observations in a finite energy range,
and if the spectra have different shapes, 
it is obvious that the differential flux at one single energy 
is not sufficient to completely describe the spectrum 
and to determine whether the source can be resolved.

The Fermi--LAT observations, as discussed in section~\ref{sec:fermi}
show that, excluding pulsars, most gamma--ray sources have curved spectra that
can be reasonably well described by the log--parabola form with critical energies
that span a wide interval, and one can assume that this continues to be true
also in the TeV and multi--TeV range.
The Fermi--LAT telescope, that is mostly sensitive
in the sub--TeV energy range, can then resolve sources with low $E^*$
that are not observable for HESS,
and conversely the HESS telescope can resolve sources with high $E^*$
that are not identified by Fermi--LAT.

\section{Milky Way source spectrum}
\label{sec:milky_way_spectrum}

\begin{figure}
\begin{center}
\includegraphics[width=10.5cm]{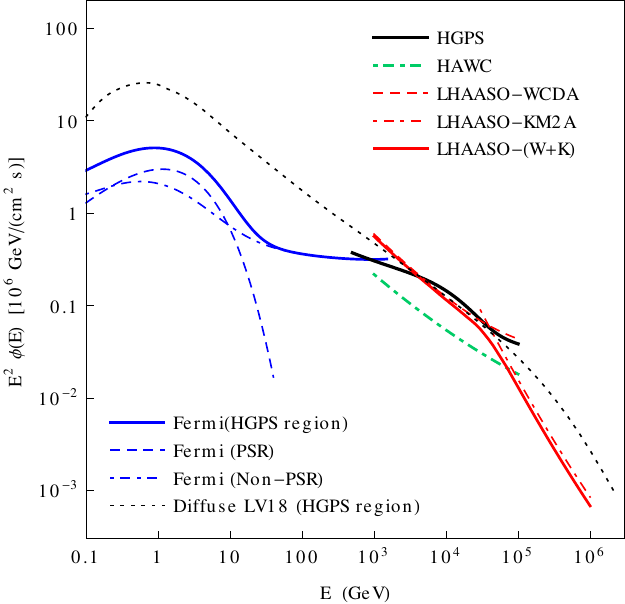}
\end{center}
\caption {\footnotesize
Plot of the cumulative spectra obtained by summing the fits to all 
sources in different gamma--ray catalogs: Fermi-LAT 
HESS-HGPS, HAWC, and LHAASO. Objects identified as extragalactic have been excluded.
For Fermi--LAT, only the sources in the sky region of the HGPS survey
have been included, in addition the contributions of pulsars,
and of all sources excluding pulsars are also shown separately.
For LHAASO the fluxes obtained by summing sources observed in the
WCDA and KM2A arrays are shown separately,
and an additional line shows the spectrum obtained by summing the combined fits
of sources observed by both arrays.
The black dotted curve shows the diffuse gamma ray flux
calculated in reference\cite{Lipari:2018gzn} for the factorized model and
integrated in the HGPS sky region.
\label{fig:all_spectra}}
\end{figure}

In Fig.~\ref{fig:all_spectra} are plotted the
cumulative spectra calculated by summing the fits to all
Galactic sources listed in the three high energy catalogs by HESS, HAWC and LHAASO,
and the cumulative spectrum obtained by summing the Fermi--LAT sources
that are observed in the same sky region of the HGPS catalog
\{$|b| < 3^\circ$, $ -110^\circ \le \ell \le +65^\circ$\},
excluding objects identified as extragalactic.
Cumulative spectra of all 4FGL sources 
classified as Galactic and of those in the
larger disk region ($|b| < 5^\circ$ and all longitudes) have been already discussed and
shown in Figs.~\ref{fig:galactic_id_spectra} and~\ref{fig:disk_spectra}, respectively.
All these three 4FGL spectra have, to a good approximation,
the same shape, indicating that the differences between 
the populations of Galactic gamma-ray sources observable in different directions,
and therefore in different regions of the Milky Way, cannot be too large.
The motivation to include in Fig.~\ref{fig:all_spectra}
the cumulative spectrum of the sources
in the HGPS sky region is to allow an easier comparison between the
Fermi--LAT and HESS results. Examining the figure one can see,
as also discussed in the previous section,
that the two spectra connect rather smoothly, both in
shape and in absolute normalization, indicating that 
the sensitivities of the telescopes for energies around one TeV are similar.

The study of Fig.~\ref{fig:all_spectra} suggests that combining the observations
of all gamma--ray telescopes it is possible to reconstruct the shape
of the Milky Way source spectrum in a wide energy interval
from 100~MeV to several PeV's.
The spectrum has a smooth but non--trivial shape, and one can
identify several significant features.
\begin{itemize}
\item For energies below few GeV the source spectrum is 
 dominated by the contribution of pulsars,
 that emit spectra with a characteristic shape:
 a very hard power--law followed by an approximately exponential
 cutoff, with cutoff energies (see Fig.~\ref{fig:cutoff_distribution1})
 in an interval that extends only up to few~GeV.
 This results in a flux that is very hard at low energies
 (a slope $\alpha \approx 1.5$ at $E \lesssim 1$~GeV) and
 softens gradually in the energy interval 1--10~GeV,
 where the PSR sources have their cutoffs.
 The pulsar contribution to the total source flux becomes strongly
 suppressed for $E \gtrsim 20$~GeV.

\item The source spectrum exhibits a hardening around $E \sim 20$~GeV,
 where the combined contributions of gamma-ray sources belonging
 to other (non--pulsars) astrophysical classes (mostly SNR and PWN)
 overtakes the PSR component.
 For $E \lesssim 1$~TeV the total source spectrum can be reasonably
 well approximated as a power--law with a spectral index of order 2.1--2.2.

\item
 For $E \gtrsim 1$~TeV, the source spectrum is measured by
 ground--based telescopes. The HESS, HAWC and LHAASO telescopes obtain
 cumulative spectra that have shapes that are in reasonable good agreement
 with each other and with Fermi--LAT observations.
 In the energy interval 1--10~TeV the source
 spectrum undergoes a modest softening, and the slope for $E\approx 10$~TeV
 is of order 2.4--2.5.

\item 
 For $E \gtrsim 25$~TeV, where only the data of LHAASO--KM2A are available,
 a new important feature is visible in the source spectrum:
 a marked softening that develops between 10 and 100~TeV,
 with the spectral index having a step of size $\Delta \alpha \approx +1.0$.
\end{itemize}

\begin{figure}
\begin{center}
\includegraphics[width=10.5cm]{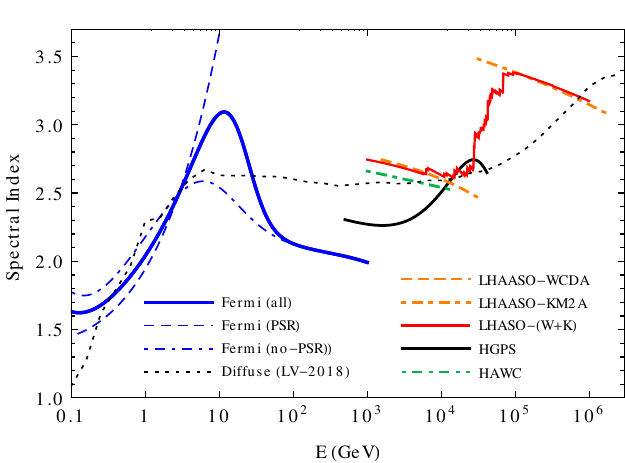}
\end{center}
\caption {\footnotesize
Spectral index of the cumulative fluxes shown in Fig.~\ref{fig:all_spectra}.
\label{fig:all_index}}
\end{figure}

The slopes of the cumulative spectra shown in Fig.~\ref{fig:all_spectra}
are plotted as a function of energy in Fig.~\ref{fig:all_index}.
The two figures obviously contain the same information, however
examining the plot of the spectral indices
one could think that the slopes obtained by different telescopes
for $E \approx 1$~TeV are not consistent,
having values of 2.00, 2.27, 2.66 and 2.75
for Fermi--LAT, HESS, HAWC and LHAASO, respectively.
The origin of this apparent disagreement can be explained noting 
the cumulative spectra have been calculated summing
fits to the measured spectra of the individual sources that are
in all cases (for HAWC and LHAASO) or for a large fraction
(for Fermi--LAT and HESS) of power--law form, while it is 
likely that a curved, softening shape would be a better description.
For a telescope that is mostly sensitive around the energy
$\tilde {E}$, where the cumulative spectrum has slope $\tilde{\alpha}$,
a power--law fit to a curved, softening spectrum
is an overestimate of the true flux at energies that are both lower
and higher than $\tilde{E}$. The reconstructed  slope
is then an overestimate 
of the true value (indicating a too soft spectrum)
for energies below $\tilde{E}$,
and is an underestimate (indicating a too hard spectrum)
for energies larger than $\tilde{E}$.

One should also note that the sum of power--law spectra of different slopes
always gives a result that is gradually hardening,
with a slope that decreases with energy
(see discussion in section~\ref{sec:cumulative_spectrum}).
This effect is clearly visible in Fig.~\ref{fig:all_index} in the energy
dependence of the slopes for the sum of the fits of sources in
the Fermi--LAT, HAWC, LHAASO--WCDA and LHAASO--KM2A catalogs.
Taking these effects into account,
the results shown in Fig.~\ref{fig:all_index} are consistent with the general
picture described above.

In Figs.~\ref{fig:all_spectra} a dashed line also shows the spectrum
of the diffuse gamma--ray flux, integrated over the HGPS sky region, calculated
with the ``factorized model'' described in \cite{Lipari:2018gzn} (LV--2018).
The slope of the calculated diffuse flux is shown in Fig.~\ref{fig:all_index}.
The LV--2018 model is constructed by assuming that the energy and angular
dependences of the diffuse flux are factorized.
The angular distribution is taken from the Fermi--LAT data
at $E \simeq 12$~GeV, where the energy distribution
is consistent with the assumption that the gamma--ray emission
is generated by the hadronic mechanism, and that cosmic rays
have spectra of the same shape in all points of the Milky Way
(as measured in the vicinity of the Earth).
The diffuse gamma--ray flux has then a spectral shape that follows
the CR all--nucleon spectrum, with a slope
of order 2.7, that softens above $E \gtrsim 100$~TeV because of the
well known ``knee'' feature.
The gamma--ray interstellar emission flux has then
a shape that is surprisingly similar to the source flux.
This is very likely only a coincidence, that however makes the separation
of the two components more difficult.

\section{Summary and outlook}
\label{sec:summary}
In this paper we have used recent observations from space and ground--based
gamma--ray telescopes to obtain an estimate of the shape of the Milky Way source spectrum,
(i.e.  the spectrum of the ensemble of all gamma--ray sources in the Galaxy)
in a very broad energy range from less than 0.1~GeV to more than 1~PeV.

The simple method that we have adopted 
is to compute a cumulative spectrum which is the sum of
the fits to the fluxes of all Galactic sources listed in the existing catalogs.
For $E \lesssim 1$~TeV the identification of the Galactic sources can be done 
either by selecting the sources associated with known Galactic astrophysical objects,
or by selecting sources at small latitudes $|b|$,
where the contamination from the (isotropically distributed) extragalactic
sources is small. At higher energies ($E\gtrsim 1$~TeV) most of the gamma--rays emitted
by extragalactic sources are absorbed during propagation, and only
a few nearby extragalactic objects can be observed. Excluding these objects,
the catalogs contain only  Galactic sources,
with very little contamination.

The simple procedure of summing the fluxes of the resolved sources
without taking into account for their distances and without correcting
for the contribution of unresolved objects has obvious limitations.
The most important one is that it does not allow,
without introducing additional information, to estimate the absolute normalization of the
Galactic emission, and thus the total luminosity of the Milky Way sources.
Estimating the absolute normalization of the total (and unresolved)
source flux is a complex problem. A recent attempt is
presented in \cite{Lipari:2024pzo}.

The method of summing the fluxes of resolved objects could in principle
lead to an incorrect estimate of the shape of the total source spectrum, because
the properties of the gamma--ray emission from nearby or bright sources
could be different from those of distant or faint sources.
It is certainly necessary to study these issues in depth, but
the tests we have performed with the available data
comparing the spectra of sources in different regions of the sky
have not revealed large differences and support the
approximate validity of the method.
Accordingly the spectral shape obtained with the simple method presented
here can be considered as a reasonable first approximation to the
shape of the total Galactic spectrum.

The reconstructed spectral shape
can be a template for the unresolved source flux
and used for the identification of the background to the 
diffuse gamma--ray flux generated by interstellar emission,
and  also  a template for the gamma--ray emission from external galaxies.

The shape of the total Galactic source spectrum obtained from our study
shows some interesting features which have been listed and discussed
in the previous section. The spectrum has a moderate softening
for energies between 1 and 10~TeV, and a larger one between 10 and 100~TeV
revealed by the LHAASO observations. The presence of this high energy
softening implies that the flux from unresolved sources 
is smaller than previously estimated, and the measurement
of the flux due to interstellar emission becomes easier.

Perhaps the most important result of our study is that the
smooth shape of the total source spectrum is formed 
by  combining  the spectra of individual sources
that have a variety of different shapes.
Most previous studies of gamma--ray source populations
have been constructed using the
approximation that all sources, in the energy range of interest,
emit spectra of the same shape. This assumption now needs to be revised.

Indeed, the finding that the spectral shape of the sum
of the fluxes of all (or a large number of) sources is quite different
from the (many) possible shapes of the individual objects
is a result that has profound implications for the modeling of the sources,
and possibly also for our understanding of the formation of the Galactic
cosmic ray spectra.

Questions of fundamental importance which we have not addressed in this paper
concern the nature of the Galactic gamma--ray sources, and the
mechanisms that generate their emissions. To answer these questions
detailed studies of a sufficiently large number of individual objects
in different astrophysical classes are needed,
but also source population studies, such as the present one,
which measure the emission of ensembles of sources can play an important role.
Our preliminary investigations suggest that several different classes of objects
are likely to give significant contributions.
More data are needed to fully understand the problem.

Another question of fundamental importance, not addressed
in this paper, is the relative importance of the hadronic
and leptonic mechanisms in generating the gamma--ray emission from
the Galactic sources. This problem remains open, and will be
discussed in more detail elsewhere. Future measurements of
high energy Galactic neutrinos will provide very important information
to obtain a robust solution.

\vspace{0.35 cm}
\noindent {\bf Acknowledgments} \\
We are grateful to Cao Zhen, Felix Aharonian,
Dmitriy Khangoulian, You Zhi Yong, Xi Shao Qian and Zha Min for useful discussions.

\clearpage

%%%%%%%%%%%%%%%%%%%%%%%%%%%%%%%%%%%%%%%%%%%%%%%%%%%%%%%%%%%%%%%%%%%%%%%%%%%%%%%%%%%%%%
%% Figures
%%%%%%%%%%%%%%%%%%%%%%%%%%%%%%%%%%%%%%%%%%%%%%%%%%%%%%%%%%%%%%%%%%%%%%%%%%%%%%%%%%%%%%

%%%%%%%%%%%%%%%%%%%%%%%%%%%%%%%%%%%%%%%%%%%%%%%%%%%%%%%%%%%%%%%%%%%%%%%%%%%%%%%%%%%%%%%%%%%

%%%%%%%%%%%%%%%%%%%%%%%%%%%%%%%%%%%%%%%%%%%%%%%%%%%%%%%%%%%%%%%%%%%%%%%%%%%%%%%%%%%%%%

%%%%%%%%%%%%%%%%%%%%%%%%%%%%%%%%%%%%%%%%%%%%%%%%%%%%%%%%%%%%%%%%%%%%%%%%%%%%%%%%%%%%%%

\clearpage

\clearpage

%%%%%%%%%%%%%%%%%%%%%%%%%%%%%%%%%%%%%%%%%%%%%%%%%%%%%%%%%%%%%%%%%%%%%%%%%%%%

\clearpage

%%%%%%%%%%%%%%%%%%%%%%%%%%%%%%%%%%%%%%%%%%%%%%%%%%%%%%%%%%%%%%%%%%%%%%%%%%%%%%%%%%%%%%%%

%%%%%%%%%%%%%%%%%%%%%%%%%%%%%%%%%%%%%%%%%%%%%%%%%%%%%%%%%%%%%%%%%%%%%%%%%%%%%%%%%%%%%%

\clearpage

%%%%%%%%%%%%%%%%%%%%%%%%%%%%%%%%%%%%%%%%%%%%%%%%%%%%%%%%%%%%%%%%%%%%%%%%%%%%%%%%%%%%

\clearpage

%%%%%%%%%%%%%%%%%%%%%%%%%%%%%%%%%%%%%%%%%%%%%%%%%%%%%%%%%%%%%%%%%%%%%%%%%%%%

%******************************************************************************

\clearpage

%%%%%%%%%%%%%%%%%%%%%%%%%%%%%%%%%%%%%%%%%%%%%%%%%%%%%%%%%%%%%%%%%%%%%%%%%%%

\clearpage

%%%%%%%%%%%%%%%%%%%%%%%%%%%%%%%%%%%%%%%%%%%%%%%%%%%%%%%%%%%%%%%%%%%%%%%%%%%%

\clearpage

%%%%%%%%%%%%%%%%%%%%%%%%%%%%%%%%%%%%%%%%%%%
% Galactic spectra of individual sources
%%%%%%%%%%%%%%%%%%%%%%%%%%%%%%%%%%%%%%%%%%% 

%%%%%%%%%%%%%%%%%%%%%%%%%%%%%%%%%%%%%%%%%%%%%%%%%%%%%%%%%%%%%%%%%%

\clearpage 

\clearpage

\clearpage


\begin{thebibliography}{100}

%\cite{Fermi-LAT:2019yla}
\bibitem{Fermi-LAT:2019yla}
S.~Abdollahi \textit{et al.} [Fermi-LAT],
``Fermi Large Area Telescope Fourth Source Catalog,''
Astrophys. J. Suppl. \textbf{247}, no.1, 33 (2020)
doi:10.3847/1538-4365/ab6bcb
[arXiv:1902.10045 [astro-ph.HE]].
%1257 citations counted in INSPIRE as of 02 Dec 2024


\bibitem{Fermi-LAT-4FGL-dr4}
S.~Abdollahi \textit{et al.} [Fermi-LAT],
``Incremental Fermi Large Area Telescope Fourth Source Catalog,''
Astrophys. J. Supp. \textbf{260}, no.2, 53 (2022)
doi:10.3847/1538-4365/ac6751
[arXiv:2201.11184 [astro-ph.HE]].
%301 citations counted in INSPIRE as of 02 Dec 2024


%\cite{HESS:2018pbp}
\bibitem{HESS:2018pbp}
H.~Abdalla \textit{et al.} [HESS],
``The H.E.S.S. Galactic plane survey,''
Astron. Astrophys. \textbf{612}, A1 (2018)
doi:10.1051/0004-6361/201732098
[arXiv:1804.02432 [astro-ph.HE]].
%405 citations counted in INSPIRE as of 02 Dec 2024


%\cite{HESS:2018pbp}
\bibitem{HAWC:2020hrt}
A.~Albert \textit{et al.} [HAWC],
``3HWC: The Third HAWC Catalog of Very-High-Energy Gamma-ray Sources,''
Astrophys. J. \textbf{905}, no.1, 76 (2020)
doi:10.3847/1538-4357/abc2d8
[arXiv:2007.08582 [astro-ph.HE]].
%175 citations counted in INSPIRE as of 02 Dec 2024


%\cite{LHAASO:2023rpg}
\bibitem{LHAASO:2023rpg}
Z.~Cao \textit{et al.} [LHAASO],
``The First LHAASO Catalog of Gamma-Ray Sources,''
Astrophys. J. Suppl. \textbf{271}, no.1, 25 (2024)
doi:10.3847/1538-4365/acfd29
[arXiv:2305.17030 [astro-ph.HE]].
%148 citations counted in INSPIRE as of 02 Dec 2024

%\cite{Fermi-LAT:2014ryh}
\bibitem{Fermi-LAT:2014ryh}
M.~Ackermann \textit{et al.} [Fermi-LAT],
``The spectrum of isotropic diffuse gamma-ray emission between 100 MeV and 820 GeV,''
Astrophys. J. \textbf{799}, 86 (2015)
doi:10.1088/0004-637X/799/1/86
[arXiv:1410.3696 [astro-ph.HE]].
%845 citations counted in INSPIRE as of 11 Feb 2025

%\cite{Fermi-LAT:2016zaq}
\bibitem{Fermi-LAT:2016zaq}
F.~Acero \textit{et al.} [Fermi-LAT],
``Development of the Model of Galactic Interstellar Emission for Standard Point-Source Analysis of Fermi Large Area Telescope Data,''
Astrophys. J. Suppl. \textbf{223}, no.2, 26 (2016)
doi:10.3847/0067-0049/223/2/26
[arXiv:1602.07246 [astro-ph.HE]].
%427 citations counted in INSPIRE as of 02 Dec 2024


%\cite{TibetASgamma:2021tpz}
\bibitem{TibetASgamma:2021tpz}
M.~Amenomori \textit{et al.} [Tibet ASgamma],
``First Detection of sub-PeV Diffuse Gamma Rays from the Galactic Disk: Evidence for Ubiquitous Galactic Cosmic Rays beyond PeV Energies,''
Phys. Rev. Lett. \textbf{126}, no.14, 141101 (2021)
doi:10.1103/PhysRevLett.126.141101
[arXiv:2104.05181 [astro-ph.HE]].
%201 citations counted in INSPIRE as of 02 Dec 2024

%\cite{LHAASO:2023gne}
\bibitem{LHAASO:2023gne}
Z.~Cao \textit{et al.} [LHAASO],
``Measurement of Ultra-High-Energy Diffuse Gamma-Ray Emission of the Galactic Plane from 10~TeV to 1~PeV with LHAASO-KM2A,''
Phys. Rev. Lett. \textbf{131}, no.15, 151001 (2023)
doi:10.1103/PhysRevLett.131.151001
[arXiv:2305.05372 [astro-ph.HE]].
%83 citations counted in INSPIRE as of 02 Dec 2024


%\cite{HAWC:2023wdq}
\bibitem{HAWC:2023wdq}
R.~Alfaro \textit{et al.} [HAWC],
``Galactic Gamma-Ray Diffuse Emission at TeV Energies with HAWC Data,''
Astrophys. J. \textbf{961}, no.1, 104 (2024)
doi:10.3847/1538-4357/ad00b6
[arXiv:2310.09117 [astro-ph.HE]].
%6 citations counted in INSPIRE as of 02 Dec 2024


%\cite{Lipari:2018gzn}
\bibitem{Lipari:2018gzn}
P.~Lipari and S.~Vernetto,
``Diffuse Galactic gamma ray flux at very high energy'' (LV--2018),
Phys. Rev. D \textbf{98}, no.4, 043003 (2018)
doi:10.1103/PhysRevD.98.043003
[arXiv:1804.10116 [astro-ph.HE]].
%93 citations counted in INSPIRE as of 02 Dec 2024

%\cite{Steppa:2020qwe}
\bibitem{Steppa:2020qwe}
C.~Steppa and K.~Egberts,
``Modelling the Galactic very-high-energy $\gamma$-ray source population,''
Astron. Astrophys. \textbf{643}, A137 (2020)
doi:10.1051/0004-6361/202038172
[arXiv:2010.03305 [astro-ph.HE]].
%30 citations counted in INSPIRE as of 02 Dec 2024

%\cite{Cataldo:2020qla}
\bibitem{Cataldo:2020qla}
M.~Cataldo, G.~Pagliaroli, V.~Vecchiotti and F.~L.~Villante,
``The TeV Gamma-Ray Luminosity of the Milky Way and the Contribution of H.E.S.S. Unresolved Sources to Very High Energy Diffuse Emission,''
Astrophys. J. \textbf{904}, no.2, 85 (2020)
doi:10.3847/1538-4357/abc0ee
[arXiv:2006.04106 [astro-ph.HE]].
%20 citations counted in INSPIRE as of 02 Dec 2024

%\cite{Luque:2022buq}
\bibitem{Luque:2022buq}
P.~D.~Luque, D.~Gaggero, D.~Grasso, O.~Fornieri, K.~Egberts, C.~Steppa and C.~Evoli,
%``Galactic diffuse gamma rays meet the PeV frontier,''
Astron. Astrophys. \textbf{672}, A58 (2023)
doi:10.1051/0004-6361/202243714
[arXiv:2203.15759 [astro-ph.HE]].
%48 citations counted in INSPIRE as of 11 Feb 2025


%\cite{IceCube:2013low}
\bibitem{IceCube:2013low}
M.~G.~Aartsen \textit{et al.} [IceCube],
``Evidence for High-Energy Extraterrestrial Neutrinos at the IceCube Detector,''
Science \textbf{342}, 1242856 (2013)
doi:10.1126/science.1242856
[arXiv:1311.5238 [astro-ph.HE]].
%1770 citations counted in INSPIRE as of 11 Feb 2025

%\cite{IceCube:2014stg}
\bibitem{IceCube:2014stg}
M.~G.~Aartsen \textit{et al.} [IceCube],
``Observation of High-Energy Astrophysical Neutrinos in Three Years of IceCube Data,''
Phys. Rev. Lett. \textbf{113}, 101101 (2014)
doi:10.1103/PhysRevLett.113.101101
[arXiv:1405.5303 [astro-ph.HE]].
%1422 citations counted in INSPIRE as of 11 Feb 2025

%\cite{IceCube:2019cia}
\bibitem{IceCube:2019cia}
M.~G.~Aartsen \textit{et al.} [IceCube],
``Time-Integrated Neutrino Source Searches with 10 Years of IceCube Data,''
Phys. Rev. Lett. \textbf{124}, no.5, 051103 (2020)
doi:10.1103/PhysRevLett.124.051103
[arXiv:1910.08488 [astro-ph.HE]].
%394 citations counted in INSPIRE as of 11 Feb 2025

%\cite{IceCube:2023ame}
\bibitem{IceCube:2023ame}
R.~Abbasi \textit{et al.} [IceCube],
``Observation of high-energy neutrinos from the Galactic plane,''
Science \textbf{380}, no.6652, adc9818 (2023)
doi:10.1126/science.adc9818
[arXiv:2307.04427 [astro-ph.HE]].
%184 citations counted in INSPIRE as of 11 Feb 2025

%\cite{Lipari:2020szc}
\bibitem{Lipari:2020szc}
P.~Lipari,
``The origin of the power\textendash{}law form of the extragalactic gamma\textendash{}ray flux,''
Astropart. Phys. \textbf{125}, 102507 (2021)
doi:10.1016/j.astropartphys.2020.102507
[arXiv:2001.00982 [astro-ph.HE]].
%3 citations counted in INSPIRE as of 11 Dec 2024

%\cite{Lipari:2017jou}
\bibitem{Lipari:2017jou}
P.~Lipari,
%``Spectral features in the cosmic ray fluxes,''
Astropart. Phys. \textbf{97}, 197-204 (2018)
doi:10.1016/j.astropartphys.2017.11.008
[arXiv:1707.02504 [astro-ph.HE]].
%19 citations counted in INSPIRE as of 11 Dec 2024

%\cite{Lipari:2024pzo}
\bibitem{Lipari:2024pzo}
P.~Lipari and S.~Vernetto,
``Resolved and unresolved Galactic gamma-ray sources,''
[arXiv:2412.08861 [astro-ph.HE]].
%1 citations counted in INSPIRE as of 10 Feb 2025

\end{thebibliography}
\end{document}